\documentclass{aa}
\pdfoutput=1
\usepackage{multirow}
\usepackage{graphicx}
\usepackage{txfonts}
\usepackage{amsmath}
\usepackage{changes}
\usepackage{natbib}
\usepackage{hyperref}
\usepackage{amssymb}
\hypersetup{
  colorlinks   = true, 
  urlcolor     = blue, 
  linkcolor    = blue, 
  citecolor   = blue 
}
\usepackage{color}
\usepackage{natbib,twoopt}
\bibpunct{(}{)}{;}{a}{}{,} 
\makeatletter
\newcommandtwoopt{\citeads}[3][][]{\href{http://adsabs.harvard.edu/abs/#3}%
{\def\hyper@linkstart##1##2{}%
\let\hyper@linkend\@empty\citealp[#1][#2]{#3}}}
\newcommandtwoopt{\citepads}[3][][]{\href{http://adsabs.harvard.edu/abs/#3}%
{\def\hyper@linkstart##1##2{}%
\let\hyper@linkend\@empty\citep[#1][#2]{#3}}}
\newcommandtwoopt{\citetads}[3][][]{\href{http://adsabs.harvard.edu/abs/#3}%
{\def\hyper@linkstart##1##2{}%
\let\hyper@linkend\@empty\citet[#1][#2]{#3}}}
\newcommandtwoopt{\citeyearads}[3][][]%
{\href{http://adsabs.harvard.edu/abs/#3}
{\def\hyper@linkstart##1##2{}%
\let\hyper@linkend\@empty\citeyear[#1][#2]{#3}}}
\makeatother
\defcitealias{jcrv2016}{Paper~I}

\begin{document}
\title{On the use of machine learning algorithms in the measurement of stellar magnetic fields}

\author{ 
     {J.C. Ram\'{\i}rez V\'elez}\inst{1}
 \and {C. Y\'a\~nez M\'arquez}\inst{2}
 \and {J.P. C\'ordova Barbosa}\inst{3} 
}
\institute{Instituto de Astronom\' ia - Universidad Nacional Aut\'onoma de M\'exico, 
           Apdo. Postal 877, 22860, Ensenada B.C.,  M\'exico,  \email{jramirez@astro.unam.mx}
 \and Laboratorio de C\'omputo Inteligente - Instituto Polit\'ecnico Nacional,  Av. Juan de Dios B\'atiz s/n, CP 07738, CDMX,
M\'exico. 
 \and CUCEA, Universidad de Guadalajara, Perif\'erico Norte 799, Los Belenes,  Zapopan Jalisco, 45100, M\'exico.
}

\offprints{jramirez@astro.unam.mx}
\date{}

\abstract
{Regression methods based in Machine Learning Algorithms (MLA) have become an important tool for
data analysis in many different disciplines.}
{In this work, we use MLA in an astrophysical context; our goal is to measure the mean longitudinal magnetic field in stars
($H_{\rm eff}$) from polarized spectra of high resolution, through the inversion of the so-called multi-line profiles.}
{Using synthetic data, we tested the performance of our technique considering different noise levels:
In an ideal scenario of noise-free multi-line profiles, the inversion results are excellent; however, the
accuracy of the inversions  diminish considerably when noise is taken into account. In consequence, we propose a
data pre-process in order to reduce the noise impact, which consists in a  denoising profile process combined  
with an iterative inversion methodology.}
{Applying this data pre-process, we have found a considerable improvement of the inversions results, 
allowing to estimate the errors associated to the measurements of stellar  magnetic fields at different noise levels.}
{We have successfully  applied our data analysis technique to two different stars, attaining by first time the measurement of
$H_{\rm eff}$ from multi-line profiles beyond the condition of line autosimilarity  assumed by other techniques. }


\titlerunning{Machine learning algorithms applied to the measurement of $H_{\rm eff}$.}
\authorrunning{Ram\' irez V\'elez  et al.\ \ }
\maketitle

\section{Introduction}

Magnetic fields are a key ingredient in the study of the stellar evolution 
and yet many of their aspects 
are still unkown, among  other reasons, due to the difficulty to 
detect, measure and map (whenever this is possible) 
surface magnetic fields in  stars  other than the Sun. 

Nonetheless, a lot of progress has been made in the recent years in the study
and observation of stellar magnetic fields, and  some facts are now 
well-stablished. For example, non chemically peculiar massive stars
can host stable strong dipolar configurations and the mean 
longitudinal magnetic  fields ($H_{\rm eff}$) for 
these stars vary from tens to thousands of gauss
\cite[e.g][]{donati2009, wade2016, grunhut2017}; 
or that the solar type stars during the main sequence phase, 
harbor more complex magnetic geometries and exhibit
much weaker effective fields ($H_{\rm eff}$), by at least one order of magnitude, 
namely, in the order of tens of gauss or less \citep[e.g.][]{petit2008,marsden2014}.
These two examples are of interest for this work from a data analysis point of view, 
because their results were obtained using the so-called multi-lines 
approach. This means that to retrieve a signal in the circular polarised 
Stokes parameter ($V$),  in order to detect and to subsequently measure the magnetic field, 
it is necessary to co-add as many as possible individual spectral lines. The line 
addition permits to boost the signal-to-ratio value in the  resulting ``mean'' 
circular profile. Otherwise, the individual $V$ profiles are in general 
systematically bellow the noise level.

The underlying idea of grouping many lines toghether into a single one is that the
noise addition is incoherent while the  addition of the $V$ profiles is coherent 
\citep{semel1996}, i.e., the more individual profiles are co-added,  the better 
signal to noise ratio is retrieved in the mean profile. 
Typically, an ad-hoc selection of unblended lines is performed establishing a linear 
relation between then mean circular profile and the mean longitudinal magnetic field.
The  number of lines to combine depends on the spectral type of the star and on the 
instrumental resolution. In the case of modern high resolution spectroplarimeters such 
as HARPS, ESPaDOnS or NARVAL, the total number can reach up to 
ten thousand lines, or more,   for late-type stars.

Different techniques have been proposed to perform the line addition, and they can be
roughly divided into two groups: 1) the line addition is based on the 
assumption that all the  lines are autosimilars \citep[e.g.][]{donati1997,wade2000,
kochu2010} 
or apply deblending process to extend the range of applicability of the technique
\citep{sennhauser2009, sennhauser2010} and 2)  the 
line autosimilarity assumption is not necessar \citep[e.g.][]{semel2006,semel2009,jcrv2010}.
The mean profiles obtained in the former approach are called  Least-Square-Deconvolved 
(LSD) profiles after the work of \citet{donati1997},   and they require the use of unblended 
lines, while the mean profiles issued from the latter approach are called Multi-Zeeman-Signatures (MZS) 
after  the work of \citet{semel2009}, and in this case the mean profiles are established
including blended and unblended lines.

For the LSD profiles, the measurement of the strength of the mean longitudinal
magnetic fields can be done through the weak field approximation 
\citep[WFA, see e.g.][]{jefferies1989}, or alternatively, with the centre-of-gravity 
method \citep[COG, see e.g.][]{rees1979} through the first-order 
moment of the mean circular profile \citep{mathys1989}. 
The employment of the WFA and of the COG methods  are of great 
utility because they allow to measure the effective magnetic field withouth having 
to recourse to radiative transfer calculations. Nowadays, the COG method is the technique
used by excellence to determine $H_{\rm eff}$ from the LSD profiles.
Nevertheless, both approaches have limitations due
to assumption of self-similarity of the lines, 
having to fulfill simoultaneously both contidions: 
the use of unblended lines and the measurement of
\emph{weak} magnetic fields. The range of application in the data analysis
is thus difficult to determine precisely. 
For example, concerning the second condition related to the strength of the magnetic fields, 
\cite{kochu2010} stated a maximum of 2 kG as limit for the analysis of circular
polarised LSD profiles; however, as noticed by
\cite{carroll2014} the limit of the WFA of local profiles, i.e. to reach
the Zeeman saturation regime in individual lines,  is around one kG
for lines in the optical wavelength range. In any case, to bypass the inherent limitations  
associated to the use of the LSD profiles and to disposse of a more general approach, 
in this work we continue developping a method based in the analysis of the MZS profiles.

In a previous work, \citet[][hereafter Paper I]{jcrv2016}, we presented a
new inversion code for the measurement of stellar magnetic fields using a 
complete radiative transfer approach. We used as case of study  a solar type 
star ($T_{\rm eff} = 5\,500\,K$), to show that the MZSs can be used to 
correctly retrieve the stellar longitudinal magnetic field in cool stars. 
The proposed methodology illustrates that it is feasible to infer $H_{\rm eff}$, 
in the context of the multi-lines approach, beyond the assumption of 
lines autosimiliraity, i.e, beyond the employment of the COG method (or the WFA).

The inversion code used in \citetalias{jcrv2016} was based on a look-up table, i.e, 
the higher the number of stellar MZSs contained in the table the better were the results
obtained. This fact makes of our technique a very CPU-time consuming method. 
In this work we incorporate the use of machine learning algorithms to remarkably 
decrease the number of synthetic stellar spectra required to properly determine 
$H_{\rm eff}$.

In section II, we show the performance of different machine learning algorithms for the 
inversion of ideal noiseless MZSs, to subsequently show that the effectivity of the
regression alrogrithms is strongly affected when noise is included, even for very small 
noise levels. In section III we incorporate pre-process data analysis  to reduce 
the noise effect when infering $H_{\rm eff}$. In section IV we apply our
method to two real cases and compare our results to previous ones published in others 
studies, to finally drew some general conclusions in section V.

\section{The machine learning strategy}

In the stellar pyhisics domain, the number of published studies using 
analysis methods based in machine learning algorithms has been growing,
specially in the last years, either for classification of stars
\citep[e.g.][]{richards2011, armstrong2016}
or in the search of exoplanets \citep[e.g.][]{davies2016}, or in the 
determination of stellar fundamental parameters 
\citep[e.g.][]{bellinger2016, verma2016, angelou2017}, among many others.
In this work we will use the machine learning approach in the 
context of the study of stellar magnetic fields.

As previously mentioned, in \citetalias{jcrv2016} we performed inversions of MZSs
using a code based in a look-up table strategy, meaning that given a MZSs to invert 
we find as solution the closest-norm MZSs in the table.  The MZSs 
were constructed 
adopting the eccentric tilted dipole model \citep{stift1975}. In this model, we 
considered as free: 1) all the parameters that describe the magnetic geometry of 
the dipolar configuration (the three Eulerian angles $\alpha, \beta, \gamma$, and the 
inclination angle ($i$) of the stellar rotation axis respect to the line of sigth),  
2) the position of the magnetic dipole inside of the star given by two coordinates ($X_2,X_3$),
and 3) the strength of the dipolar moment ($m$), 
leading to a total of 7 free parameters. It turned out that in order to obtain good
enough inversion results, it was required to calcute $7\,500$ stellar spectra 
models. For each synthetic spectrum we randomly varied the 7 free parameters,
covering a spectral range from $350$ to $1\,000$ nm,  giving a total  
close to $350\,000$ wavelength points per spectra. This huge number of 
thousands of stellar spectra, required to shape the table,
demands a lot of computer calculation time which damps the employment of
our techinque for the analysis of big databases,  such as 
{\em Polarbase}\footnote{http://polarbase.irap.omp.eu/}
\citep{petit2014}, where  hundreds of observed spectra of cool stars can be found.

To reduce the required number of stellar spectra to invert the MZSs, 
in this work we implemented 3 different regression methods based in machine 
learning algorithms.  All these three regression models belong to the 
supervised paradigm and are included in the \emph{Scikit-learn}
software package \citep{scikit} of the Python language. 
The employed regressions are: 1) The Bayesian Ridge Linear Regression (BR),
2) The Support Vector Machine (SVM), and 3) The Multi Layer Perceptron Artificial Neuronal 
Network (ANN). We next briefvly describe each one.


Regression analysis is a valuable mechanism to model and analyse data, constituting a form of 
predictive modelling technique used to explore the relationship between two (or more) variables: 
an independent one, the input (also known as feature or predictor), and a dependent one, the target.
Its goal is to  fit a curve to the data points in such a form that the sum of 
differences between the data points and the curve is minimal \citep[e.g.][]{tan2005}. 
We next briefly describe each of the regressor models employed in this work.

\subsection{Bayesian Ridge}

Linear Regression is one of the most frequently used modeling techniques:
It establishes  a linear relationship between the 
dependent variable  ($y$) and one or more independent variables ($\vec{x}$):

\begin{equation}
 y (\vec{x},\vec{w})=w_0 + w_1x_1 + w_2x_2 + ... +w_nx_n,
 \label{eq1}
\end{equation}
where the coefficients $w_i$ are known as the parameters of the 
regression model. The first coefficient $w_0$ is the so-called 
interceptor or ``bias'' of the model. In our case, the input variable 
are the MZSs and the target can be any parameter
of either the stellar atmospheric model or the magnetic model 
(as for example $T_{\rm eff}$ or $H_{\rm eff}$).

The linear  model that best fits the data is commonly calculated using the Least
Square Method and when this is the case, the performance of the
calculated model is measured by the R-square metric \citep{glantz2016}. It is important to notice 
that this kind of regression require that a linear relationship between both
independent and dependent variables exists. It is also noteworthy the fact that
outliers terribly affect the regression line and, therefore, the estimated values.

If the power of the predictor on the estimated regression equation is bigger
than one, then the regression is known as polynomial. In that case, the regression
line is not a straight line but a curve that fits to the points. A higher power on
the independent variable allows a better fit for more complex datasets.

However, when the data experiences from multicollinearity, meaning that
it has multiple, highly correlated predictors, then the Ridge Regression technique
is used. When multicollinearity occurs, the Least Squares estimates are
unbiased but the variances are large, which in turn results on a higher error
on the predicted values. To correct this error, Ridge Regression adds a bias to
the regression estimates and it solves the multicollinearity problem through a
shrinkage parameter $\lambda$ :

\begin{equation}
 min || y-\vec{x} \vec{w} ||_2^2 + \lambda ||\vec{w}||_2^2
\end{equation}

The shinkage parameter, $\lambda \ge 0$, penalizes the size of the coefficients on the 
original equation and it is the direct regulator of the amount of shrinkage, a process also
known as regularization \citep[e.g.][]{yang2015}.

The regularization of the parameters, can be done using the Bayesian Regression techinque: it
introduce diffuse priors, also known as noninformative priors, which are 
probability distributions expressing general information. Using the Bayesian approach,
the regularization of the Ridge Regression model is equivalent to 
finding a maximum solution for a Gaussian prior (the previously mentioned diffuse prior)
with precision $\lambda^{-1}$.  However, instead of setting lambda manually, it is possible to 
consider it as a random variable to be estimated from the data if the predicted output $y$
is assumed to  follow a Gaussian distribution around $\vec{x}\vec{w}$ :

\begin{equation}
 p(y | \vec{x}, \vec{w}, \alpha) = N (y | \vec{x}\vec{w}, \alpha),
\end{equation}
where $\alpha$ is a random variable and it needs to be estimated from the dataset.
A particular case of Bayesian Regression is the Bayessian Ridge Regression, where 
the prior for the parameter $\omega$ is given by a spherical Gaussian distribution, 
as in:

\begin{equation}
 p(\omega | \lambda) = N (\omega|0,\lambda^{-1} \vec{I}),
\end{equation}
where $\vec{I}$ is the identity matrix. It is during the fit of the model
that the set of parameters $\omega, \lambda, \alpha$ 
are estimated jointly \citep{carlin2008, scikit}.

\subsection{Support Vector Machines}

Support Vector Machines is a learning method used mostly to classify data and to detect
outliers, but  it can also be used for regression purposes. The SVM offers important 
characteristics in his performance, as for example high effectiveness in high 
dimensional spaces, usefulness even when the number of samples is lower than 
the number of dimensions and the memory efficency. 
In the SVM approach, data is plotted as points in a n-dimensional space
and the outcome of the algorithm will be an  optimal hyperplane that best 
categorizes the data clases  (for calssification purposes) or best predict 
the tendency of the data in order to predict future unkown data 
(regression model) \citep[e.g.][]{du2014, garcia2018}.

In the specialized literature, two main modalities are reported to perform regression 
tasks through the use of the SVM model, which are: the $\epsilon$-support vector regression ($\epsilon$-SVR) 
and the $v$-support vector regression ($\nu$-SVR), \cite{vapnik1999}. 
In this work we have used the former approach $\epsilon$-SVR:
Let $\vec{X}$ = ($\vec{x_1},\vec{x_2},..,\vec{x_m})$
be a set of observations of the same independent variable, 
and let $\vec{y}$ = ($y_1,y_2,...,y_m$) the corresponding dependent variables. Also, let $\vec{w}$ be the weight vector, 
let $b$ be a scalar, let $\Phi(\vec{x})$ be a non-linear function, let $\xi_i$  be slack variables, and
let $C>0$ be their associated parameter. If $\epsilon>0$ is the associated parameter with the 
$\epsilon$-insensitive loss function, the $\epsilon$-SVR solves the following optimization problem:

\begin{equation}
 \min _{w,b,\xi,\xi^*} \frac{1}{2} \vec{w^T}\vec{w} + 
 C \sum_{i=1}^{m} \xi_i +  C \sum_{i=1}^{m} \xi_i^*,
\end{equation}
subject to
\begin{equation}
\begin{aligned}
\vec{w^T}\Phi(\vec{x_i}) + b - y_i, \le \epsilon+\xi_i, \\
y_i -\vec{w^T}\Phi(\vec{x_i}) -b, \le \epsilon+\xi_i^*, \\
\xi_i,\xi_i^* \ge 0,i=1,...,m.
\end{aligned}
\end{equation}

Similarly, the dual problem is defined according to $\vec{\alpha_i}$:
\begin{equation}
 \min _{\alpha, \alpha^*} \frac{1}{2} (\vec{\alpha} -\vec{\alpha}^*)^T -
 Q(\vec{\alpha} -\vec{\alpha}^*)
\end{equation}
subject to 
\begin{equation}
\begin{aligned}
\vec{e^T} (\vec{\alpha} -\vec{\alpha}^*) = 0 \\
0 \le \alpha_i, \alpha_i * \le  C, \,\,   i=1,...,m
\end{aligned}
\end{equation}
where the $e^T$ is the identitity vector and $Q_{ij}$ = $K(\vec{x_i},\vec{x_j})$ $\equiv \Phi(\vec{x})^T \Phi(\vec{x})$  
is the kernel function, 
which can be linear, polynomial, sigmoid or a radial basis function, among other options. 
The dual problem is solved by the Lagrange multipliers, and the resulting function is \citep{chang2011libsvm}:

\begin{equation}
f(\vec{x}) = \sum_{i=1}^m (-\alpha_i +-\alpha_i^*) K(\vec{x_i},\vec{x_j})+b.
\end{equation}


In our case, i.e. data analysis of MZSs,  we found after  testing differnet kernel fucntions 
that it was with the linear function that we  obtained the best predictions results.

\subsection{Artificial Neuronal Network}

Artificial neural networks are inspired by physiological knowledge of the organization of the brain. 
They are structured as a set of interconnected identical units known as artificial neurons, and the 
interconnections are used to send signals from one neuron to the others, in either an enhanced or 
inhibited way. This enhancement or inhibition is obtained by adjusting connection weights. In an 
artificial neural network, a new observation causes an update of the network weights, which means 
that the network {\em learns}.

Let $x_i \in \mathbb{R}^n$ be the $i$-th observation from a set of $m$ of them (corresponding to 
the same independent variable), and let $x_{ij} \in \mathbb{R}$ be the $j$-th value of the $i$-th 
observation. From these $m$ values and a previously established threshold value $\theta$, it is 
possible to model a simple perceptron, which is an artificial (and very effective) model of 
a physiological neuron \citep{rosenblatt1958perceptron}. 
It is considered that the $j$-th value of the $i$-th observation 
(corresponding to a dendrite of a physiological neuron) has an associated weight $w_{ij} \in [0,1]$ 
when $x_{ij}$  becomes the entrance to the central part of the neuron (which corresponds to the {\em soma} 
of the physiological counterpart). The weight with value 1 is associated with the threshold value 
$\theta$.

The process that occurs in the {\em soma} of the artificial neuron, is what gives the power to the 
machine learning algorithms based on artificial neural networks. This process involves a sum of the 
products and a function of activation of the neuron. Each product is obtained by multiplying the 
value of the neuron $x_{ij}$ with its associated weight $w_{ij}$ while the activation function
$\Theta$ is typically a sigmoid function or the tanh function. The result of this process, which 
is the output of the artificial neuron for of the $i$-th observation, is shown in the following expression:

\begin{equation}
 output(i)= \Theta \, \big[ \, \big( \sum_{j=i}^n w_{ij} \,x_{ij}  \big) - \theta \, \big].
\end{equation}


The simple perceptron is the fundamental building block of virtually all models of artificial neural 
networks of the state of the art (with some singular exceptions represented by unconventional models, 
such as morphological and Alpha-Beta models). The choice of the ways of interconnecting the simple 
perceptrons gives rise to different network topologies, among which stand out, undoubtedly the models
known as {\em fully connected layered feedforward networks}.

This type of models are organized in layers: an input layer, an output layer and one or more hidden 
layers. One of its important features is no connection or feedback between neurons of the same layer,
but the connections, as the name implies, are all feedforward; that is, a neuron located in a certain
layer, only connects with one or several neurons of the adjacent layer forward \cite[e.g.][]{du2014}.

The Multi Layer Perceptron (MLP) is a fully connected layered feedforward network that is trained 
with the method known as backpropagation, whose detailed description can be consulted in 
\cite{rumelhart1985}. The MLP can be used as a classifier or as a regressor, and in this article 
we use a regressor version of the MLP known as MLPRegressor, which is trained with backpropagation 
using identitity as with the {\em identity} function as the activation function in the output layer, 
that corresponds to the output $y$ of our application \citep{scikit} .

\subsection{Regression model parameters}
In practice, the implementation of each regressor model is defined by a 
set of hyperparameters, as for example the activation function  or the value 
of the penalization coefficients, etc. The choice of each hyperparameter 
for the three regressors used in this work are all included in Appendix A.

For the choice of the different hyperparameters in each regression model
we tested the inversion effectivity by varying the values of many  hyperparameters;   
the final values that were selected (listed in Appendix A), were those that offered  
better inversion performance for each regressor.
 
Finally, for the cases of the  SVM and the ANN regressors, the associated 
learning algorithms are sensitive to the data distribution. Both regressors assume 
that the data is centred around zero and is close to a standard normal distribution.
For this reason, it is requiered to apply a standardization of the data before the 
traning is performed. We have thus applied a pre-process to the data prior to the 
training process through the {\em StandardScaler}  utility included also in the 
{\em Scikit-learn} software package.

\section{Inverting MZSs}
 
In this section we will employ the three different regression models
mentioned above to estimate the $H_{\rm eff}$ using a supervised training. The reason
for using three different models is to determine to what extent
the results depend on the choice of the model, and thus to choose 
the most adequate one.

\subsection{Inverting noiseless MZSs}

The MZSs that we will employ in this section are those previously established 
in \citetalias{jcrv2016}, where we considered the case of a cool star adopting 
from the Atlas9 grid \citep{castelli2004} an atmospheric model with 
$T_{\rm eff} = 5\,500\,K$, $\log g = 4.0$, solar metallicity, zero macro-turbulence 
($v_{\rm turb} = 0.0$) and a micro-turbulent velocity of $\xi = 2.0$\,km\,s$^{-1}$.
Besides, we considered a slow rotator case fixing $vsini = 10\,{\rm km}\,{\rm s}^{-1}$. 
For the synthesis of the Stokes profiles we employed the code {\sc Cossam} 
\citep{stift2000, stift2012}. All the atmospheric parameters were kept fixed, but 
we allowed to randomly vary the 7 parameters that describe the configuration 
the dipolar tilted eccentric model (see above). The inversions showed that none of the free 
parameters could be retrieved being only possible to infer $H_{\rm eff}$. The 
fact that none of the free parameters of the dipolar model could be retrieved 
is due to the existent degeneracy among different combinations of parameters 
that can produce the same observable: $H_{\rm eff}$ (see Fig. 10 in \citetalias{jcrv2016}). 
For this reason, the machine learning algorithms will be trained to predict 
only one target, the $H_{\rm eff}$.  We point out that our goal is to 
measure the stellar longitudinal magnetid field from spectropolarimetric snapshot 
data type,  and not to map the strength of magnetic field over the surface.

As first step, we must determine the (small enough) number of MZSs that 
we will use to train the regression models, keeping
in mind that it is required to obtain satisfactory and acceptable predictions
(inversion results).

\begin{figure}
\hspace{0.5cm}\includegraphics[width=7.5cm]{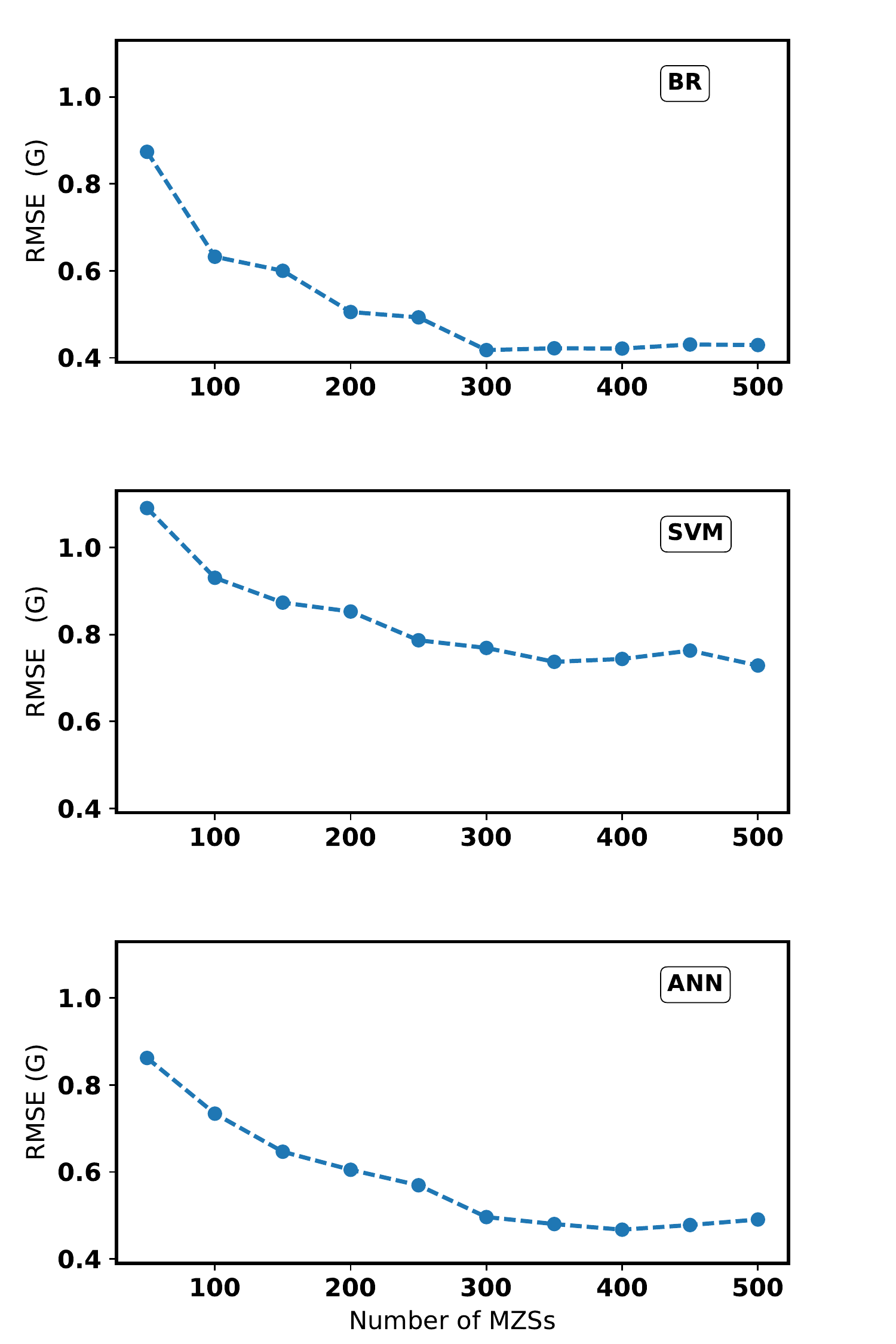}
\caption{Root mean square inversions errors (RMSE), in units of gauss,  as function of 
the number of MZSs included in the training; the inner legend in each panel
indicates the employed regression model (see text for details). The inversion sample 
consist of $1\,500$ MZSs.}
\label{fig:rmse_noiseless}
\end{figure}

\begin{figure*}
\begin{center}
\hspace{0.5cm}\includegraphics[width=14cm]{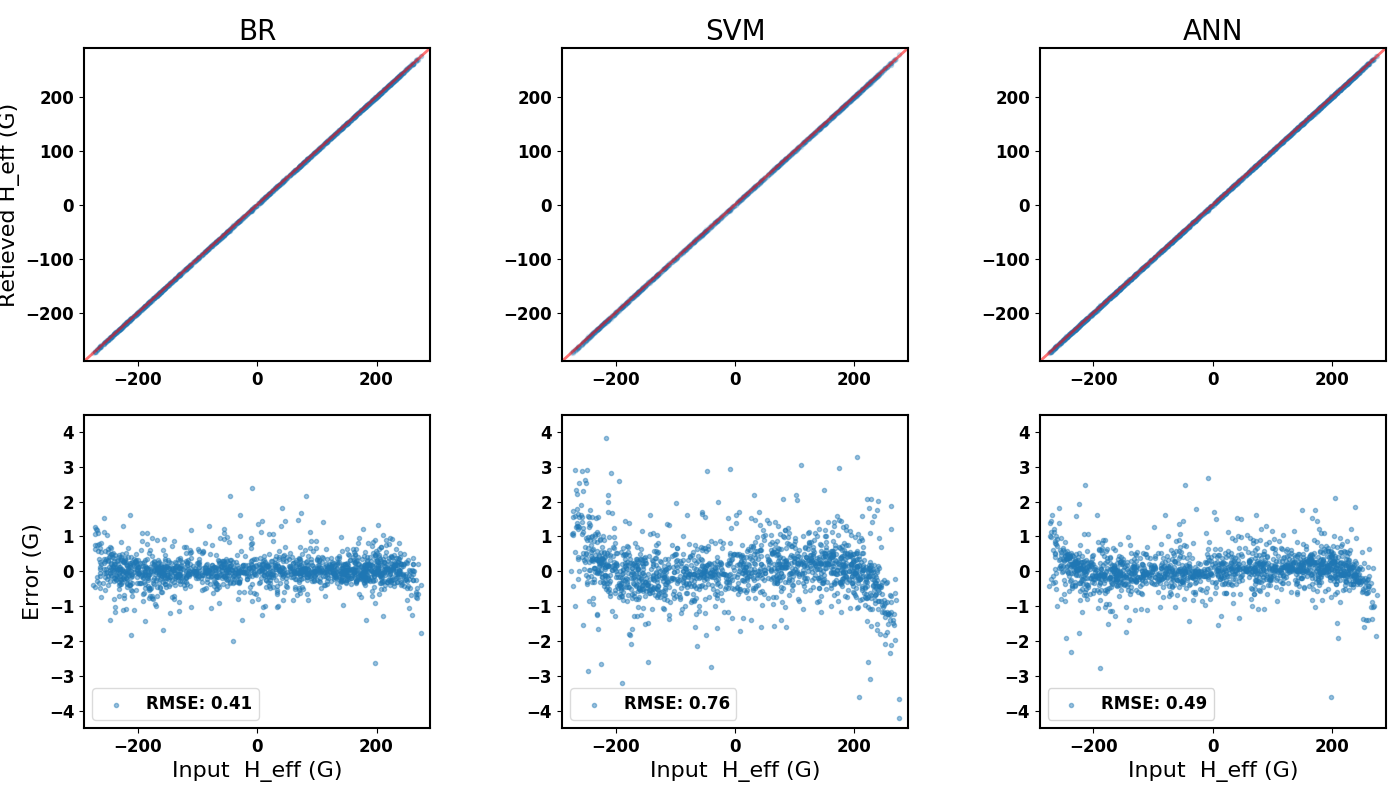}
\caption{Performance of the regression models when 300 MZSs were considered
during the training process. 
The X axis for all panels corresponds to the original 
value of $H_{\rm eff}$  for each of the $1\, 500$ MZSs. In the upper panels the Y axis 
represents the predicted values of the inversions, while in the lower panel it corresponds 
to the inversion errors in units of gauss. The red line in the upper panels represents a 
one-to-one relation.}
\label{fig:plot_inv_noiseless}
\end{center}
\end{figure*}

For this purpose, we first used only 50  MZSs to train the algorithms 
and then we tested each regression model inverting a representative set of  $1\, 500$ 
MZSs. We then repeated the exersise but increasing the number of MZSs used 
in the training process by steps of 50 until reach a maximum of 500. The performance 
for all the regression models is shown in Fig. \ref{fig:rmse_noiseless}. 
The three models showed similar results, obtaining very accurate determinations
of the $H_{\rm eff}$ even  when we considered a very small number of  MZSs in the 
training process: using 100 MZSs the obtained RMSE is inferior to 1 gauss. The general
tendency is that the errors decrease smoothly as we increase the number of  MZSs
employed in the training.  Based in these results, we have decided that training
the algorithms with 300  MZSs is sufficient for a proper determination 
of the stellar longitudinal magnetic field. Thus, hereafter all the inversions tests
will be performed using 300 MZSs in the training process. In  Fig. \ref{fig:plot_inv_noiseless} 
we show the individual inversions of the $1\,500$ MZSs for the three regression models.

The excellent results of Fig. \ref{fig:plot_inv_noiseless} show the great
utility of the employment of machine learning algorithms in the inference of 
$H_{\rm eff}$ from the MZSs. As comparison, in \citetalias{jcrv2016} we constructed  
$7\,500$ spectra models and the RMSE of those inversions was 3.95 G, while with 
the best regression model (BR) we now obtain a RMSE of 0.41 G requiring only 
300 spectra models. In other words, using this regression model the inversions accuracy
is increased by one order of magnitud, and at same time the required time
to synthetize the stellar spectra models  (number of MZSs) is reduced by a factor of 25.
It is important to remark that the requiered time for the training of the algorithms is very fast :
using a single procesor the cpu-time  requiered was of 1, 5 and 120 seconds for the
BR, SVM and ANN regressors respectively; moreover,  in the case of BR and the ANN 
the time can be reduced since the training process be parallelized.

\begin{figure*}
\begin{center}
\hspace{0.5cm}\includegraphics[width=18cm]{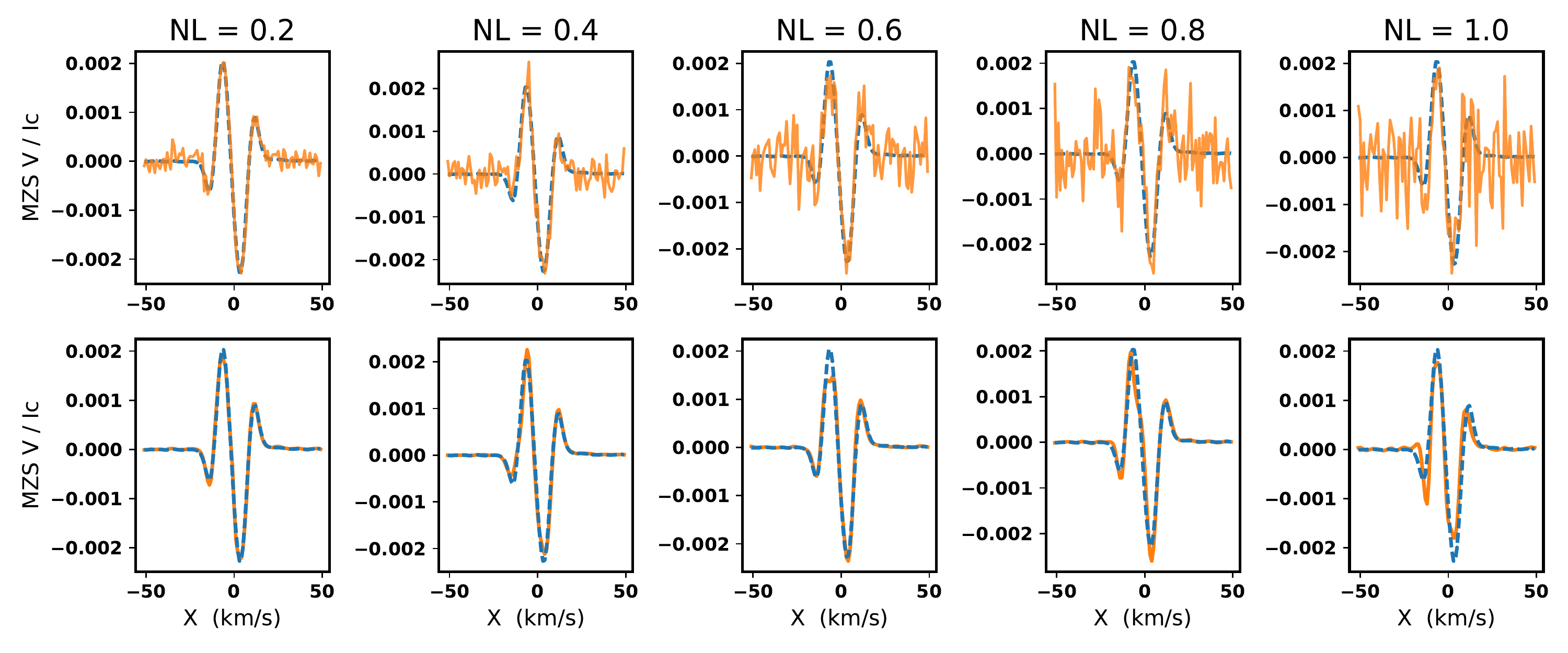}
\caption{In both, upper and lower panels, the dashed line represents the original 
(noiseless) MZSs.  The solid line in the upper panels corresponds to the noise added MZSs; 
on top of each column are indicated the levels of noise. In the lower 
panels, the solid lines represent the cleaned MZSs.}
\label{fig:cleaned_mzs}
\end{center}
\end{figure*}

\begin{figure*}
\begin{center}
\hspace{0.5cm}\includegraphics[width=18cm]{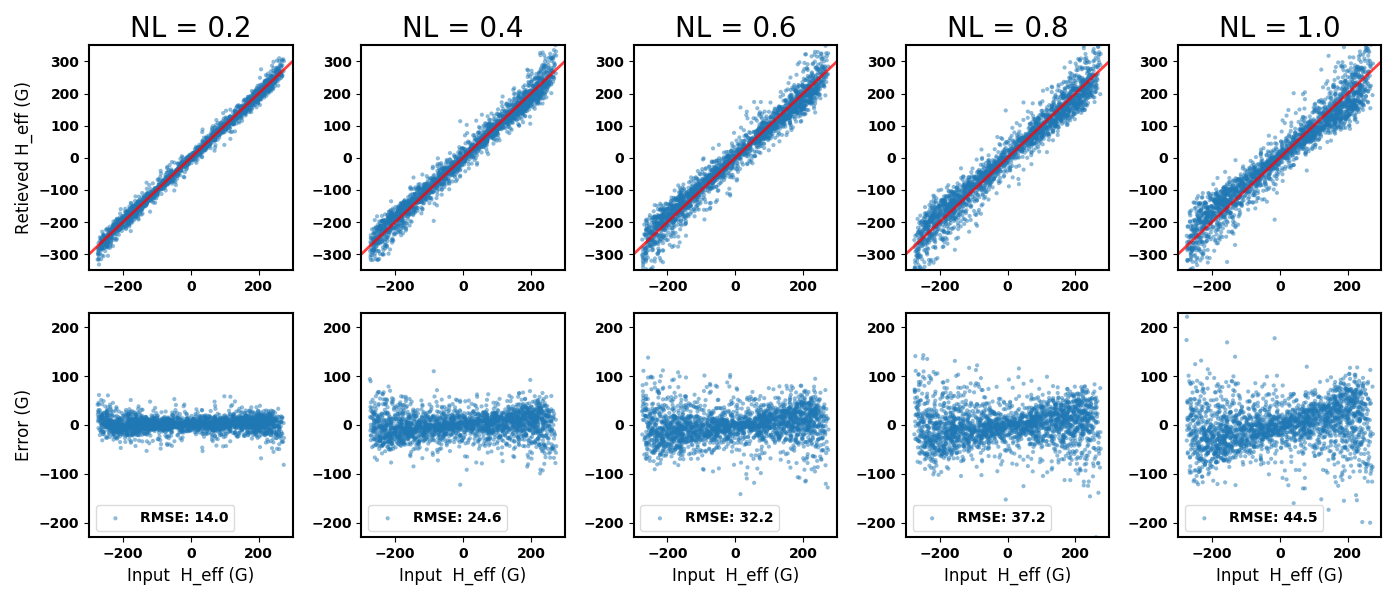}
\caption{Inversions of noise-added MZSs using the BR model.
The different noise levels (NL) added to the MZSs are indicated in top
of the upper panels. The X and Y axes are the same as in Fig. 
\ref{fig:plot_inv_noiseless}\,.}
\label{fig:br_noised}
\end{center}
\end{figure*}

Unfortunately, the effectivity of the regression models is considerably reduced when 
noise is taken to account. To illustrate this, we repeat the same exersise
of Fig. \ref{fig:plot_inv_noiseless} but this time including 5 different noise levels 
in the MZSs. The noise level,  see Fig. \ref{fig:cleaned_mzs}, 
is quantified as the standard deviation of the noise divided by the standard deviation of the noise-free MZSs.
In Fig. \ref{fig:br_noised} we 
show the inversion results only for the BR model, which was the best 
regressor for the case of ideal MZSs without noise. (We note that our
definition of noise level is consistent with the work of \cite{carroll2014}, so we can 
latter on make a proper comparison of the results.)

From the results of Fig. \ref{fig:br_noised}, it is clear that for the inversion 
of noise-affected MZSs,  the case corresponding to real data, the predictions are 
no longer good enough, even for the lowest noise level.  For these
tests, the training process was performed using noise-affected MZSs  
(with their respective noise levels), because in the contrary case 
-- training with noiseless MZSs--, the errors were even slightly higher.
Besides, it is also worth noticing that we have verified that
the number of MZSs to include in the training process does not
depend on the noise levels, i.e., to consider 300 MZSs in the 
training process is good enough for all noise levels.

Based on the obtained results when noise is included, 
we propose a two steps process to improve the efficiency of the
MZSs inversions. These two steps consist in a MZSs noise-cleaning and in an iterative 
inversion methodology.

\begin{figure*}
\begin{center}
\hspace{0.5cm}\includegraphics[width=18cm]{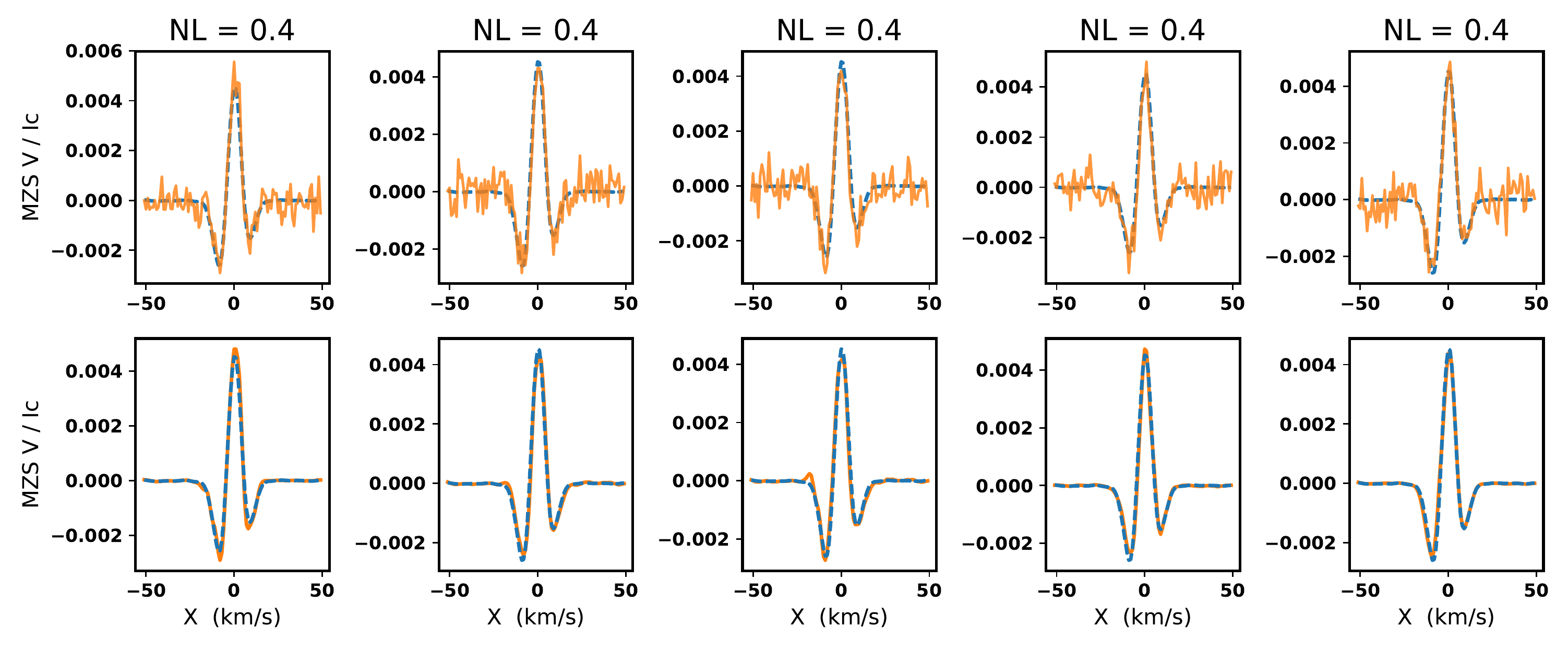}
\caption{Iterating the cleaning process considering one given MZSs and 
the same noise level. The solid and dashed lines represent the same as in Fig. 
\ref{fig:cleaned_mzs}. As shown in bottom panels, at each iteration the cleaned 
profiles show small but different deviations from the original noiseless profile.}
\label{fig:cleaned_mzs_iter}
\end{center}
\end{figure*}

\subsection{Denoising profiles with PCA}

To facilitate the analysis of observed data, it is desirable to disposse of some
noise treatment in order to give more credibility to the results.
In the case of stellar spectropolarimetric data, the establishment
of the multi-line (LSD or MZS) profiles is already a pre-processing step 
of the  data analysis. However, the multi-line profiles are also
affected with noise, which in turn degradate the accuracy of the 
inversions. Thus, we will subsequently refer as {\em noise treatment} 
to any process applied to the multi-line profiles in order to deal with 
the noise problem.

Noise treatment, in this sense, can be implemented using for example 
any of two following approaches. On the one hand, the sparsity representation 
of data samples has recently begun to  be implemented in the analysis of 
solar and stellar spectropolarimetric data with very good results 
\citep{asensio2015, carroll2014}. On the other hand, the principal 
components analysis (PCA), first introduced by \cite{rees2000} for the study of solar magnetic
fields, is a more commonly employed technique in the  analysis of both, solar and stellar 
spectropolarimetric data. Both approaches implement a sort of space dimensionality reduction 
allowing to diminish the impact of the noise in the retrieval of the model parameters.

We will employ PCA  to perform a ``noise-cleaning'' of the MZSs: In the previous section we decided 
to use 300 MZSs for the training of the regression algorithms. Applying the  
Singular Value Decomposition to this same set of MZSs we obtain in turn 300 eigenvectors. It is 
well known that the eigenvectors are a matemathical base of the original set of MZSs, such 
that any MZSs can be obtained as a linear combination of those eigenvectors \citep[e.g.][]{golub1996}.
Nevertheless, it is also possible to consider not the entire set of eigenvectors but
only the first ones, because in fact the first eigenvectors are those that contain 
most of the usefull information about the original MZSs  \cite[e.g.][]{paletou2012}. 
Accordingly, we considered only the first ten eigenvectors to perform a noise-cleaning:

\begin{equation}
 MZS_{cleaned}  \equiv \sum _{i=1} ^{10} \alpha_i  \ \ ev_i,
\end{equation}
where $\alpha_i$ denote the scalar coefficients and $ev_i$  the eigenvectors.
The coefficients can be obtained through a dot product between each eigenvector 
and the MZS.

In Fig. \ref{fig:cleaned_mzs}, we illustrate the  cleaning proces of  MZS profiles considering the 
same 5 levels of noise previously used (from 0.2 to 1.0 in steps of 0.2). The solid lines in 
the upper panels correspond the noise added profiles, while in the lower panels they 
represent the respective cleaned profiles. As comparison, the dashed lines in both, 
upper and lower panels, correspond to the original noise-free profiles. It can be noticed
that the noise cleaning process works well even for the highest noise level. 
However, as expected, the higher the noise level, the greater the difference
between the original and the cleaned MZSs are found.

We applied the described process to the full set of $1\,500$ MZSs
and repeated the same inversion test of Fig. \ref{fig:br_noised},
but this time implementing a noise cleaning prior to the inversion of the profiles. 
Surprisingly, the results are not as good as expected, being only 
slightly better to the case without cleaning process, 
as shown in Table \ref{tab1}. For this reason, we propose a complementary 
step in the data analysis  to improve  the inversion
efficiency.

\begin{table}[h]
\caption{Comparaison of the  RMSE for inversions performed considering a noise 
cleaning process of the profiles and when not. The employed regressor model was the BR. }
\label{tab1}
\centering
\begin{tabular}{c|c|c|c|c|c|c}
\hline
            & \multicolumn{5}{|c|} {Noise Level} & Cleaned \\ 
            
            &  0.2 & 0.4 & 0.6 & 0.8 & 1.0 & \\ [0.5ex]
\hline
\hline
RMSE (G)    &  14.0 & 24.6 & 32.2 & 37.2 & 44.5 & No \\  [0.75ex]

RMSE (G)    &  13.2 & 24.0 & 26.9 & 34.5 & 39.2& Yes \\  [0.75ex]
\hline
\end{tabular}
\end{table}

\subsection{Iterative inversion process}

\begin{figure*}
\begin{center}
 \includegraphics[width=16cm]{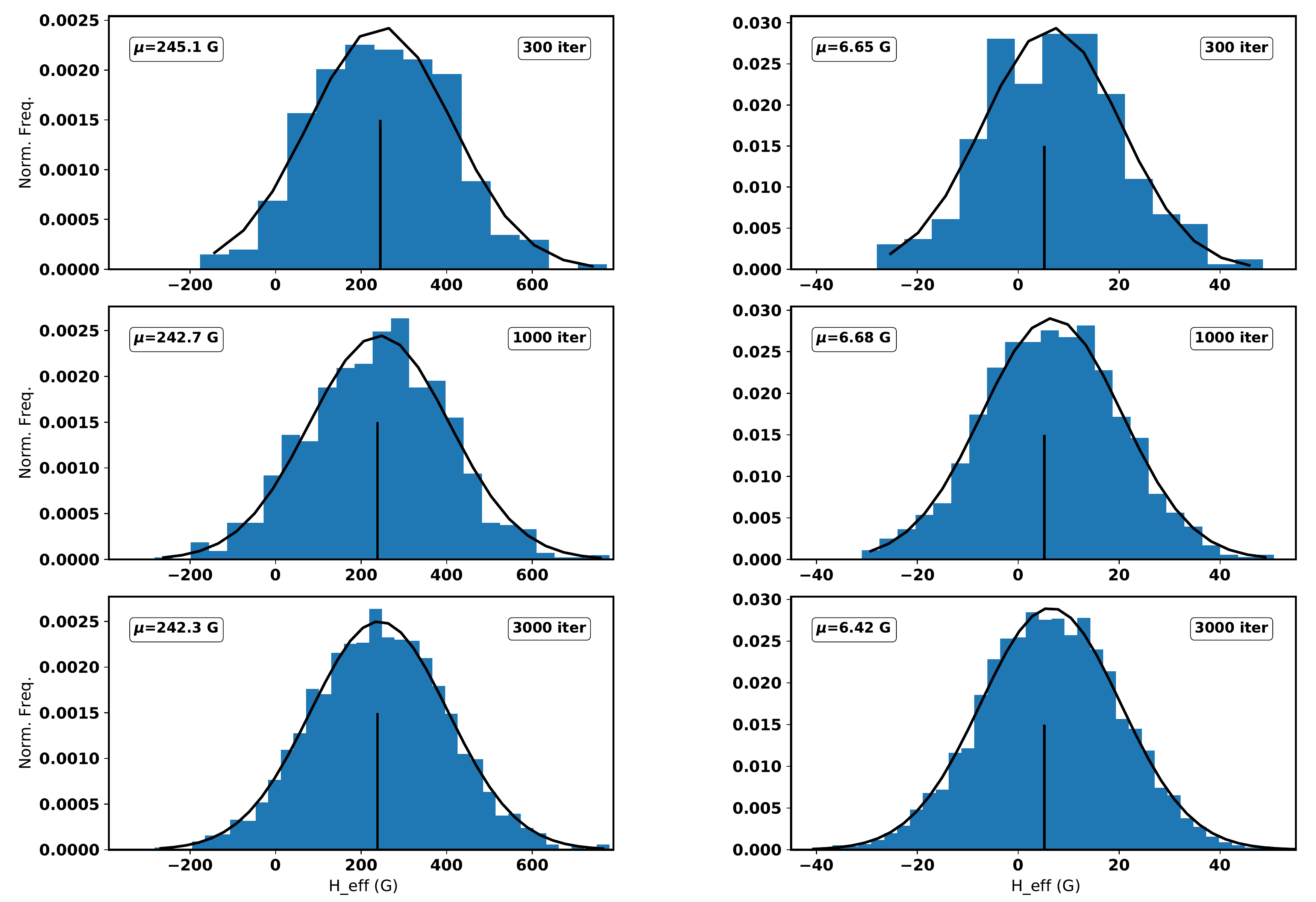}
\caption{Two examples of the iterative inversion process; the original values of $H_{\rm eff}$ are 238.2 G 
(left panels) and 5.2 G (right panels). Each histogram
shows the inversion distributions considering different numbers of iterations  
(indicated in the  upper  rigth corner). The upper left legend in 
each histogram indicates the centroid of the fitted normal Gaussian distribution, which 
would correspond to the inferred value of $H_{\rm eff}$ in each case.} 
\label{fig:histos_examples}
\end{center}
\end{figure*}

Let us consider only one given MZS from which we would like to infer $H_{\rm eff}$, 
and let us also consider a given fixed noise level, for example 0.4. 
In the previous section we showed how to perform a noise cleaning of the 
profiles. We consider now a scenario in which we could repeat this process several times.
For each iteration, the random noise will affect differently the MZS and subsequently
the noise cleaned profiles will all be  slightly differents. In Fig. \ref{fig:cleaned_mzs_iter}
we illustrate how the cleaned profiles at each iteration differ 
from the original noise-free profile. For example, the maximum amplitude for the 
bottom left cleaned profile is higher than the original one, contrary to 
the botom central cleaned profile where the maximum amplitude is inferior to the 
original one. These differences, even if very small, are enough to produce a different 
result when infering the longitudinal magnetic field. 
Nevertheless, we know that all the cleaned profiles have the same origin in the 
sense that all were established from the same noiseless data (dashed lines in Fig. \ref{fig:cleaned_mzs_iter}). 
It is thus expected that the inversions derived from these cleaned profiles will be close to the real value of
$H_{\rm eff}$, and in fact they are, showing a distribution centred around
the original value of $H_{\rm eff}$.

%
%
%
%
%
%
%
To illustrate our proposed methodology,  in Fig. \ref{fig:histos_examples} we show two examples
of iterative inversions. We have considered two MZSs, one with an  $H_{\rm eff}$
of 238.2 G (left panels) and one with a much weaker magnetic  
longitudinal field of 5.2 G (right panels). From top to bottom, the three panels show 
how the inversion distributions varies with the number of iterations: 
the larger the number of iterations performed, the more the histograms will 
approximate a normal Gaussian distribution (solid line in black). The centroid of the fitted  
Gaussian distribution, indicated with a vertical line, corresponds to the inferred 
value of $H_{\rm eff}$ in each histogram.

\begin{figure}[h]
\begin{center}
 \includegraphics[width=8cm]{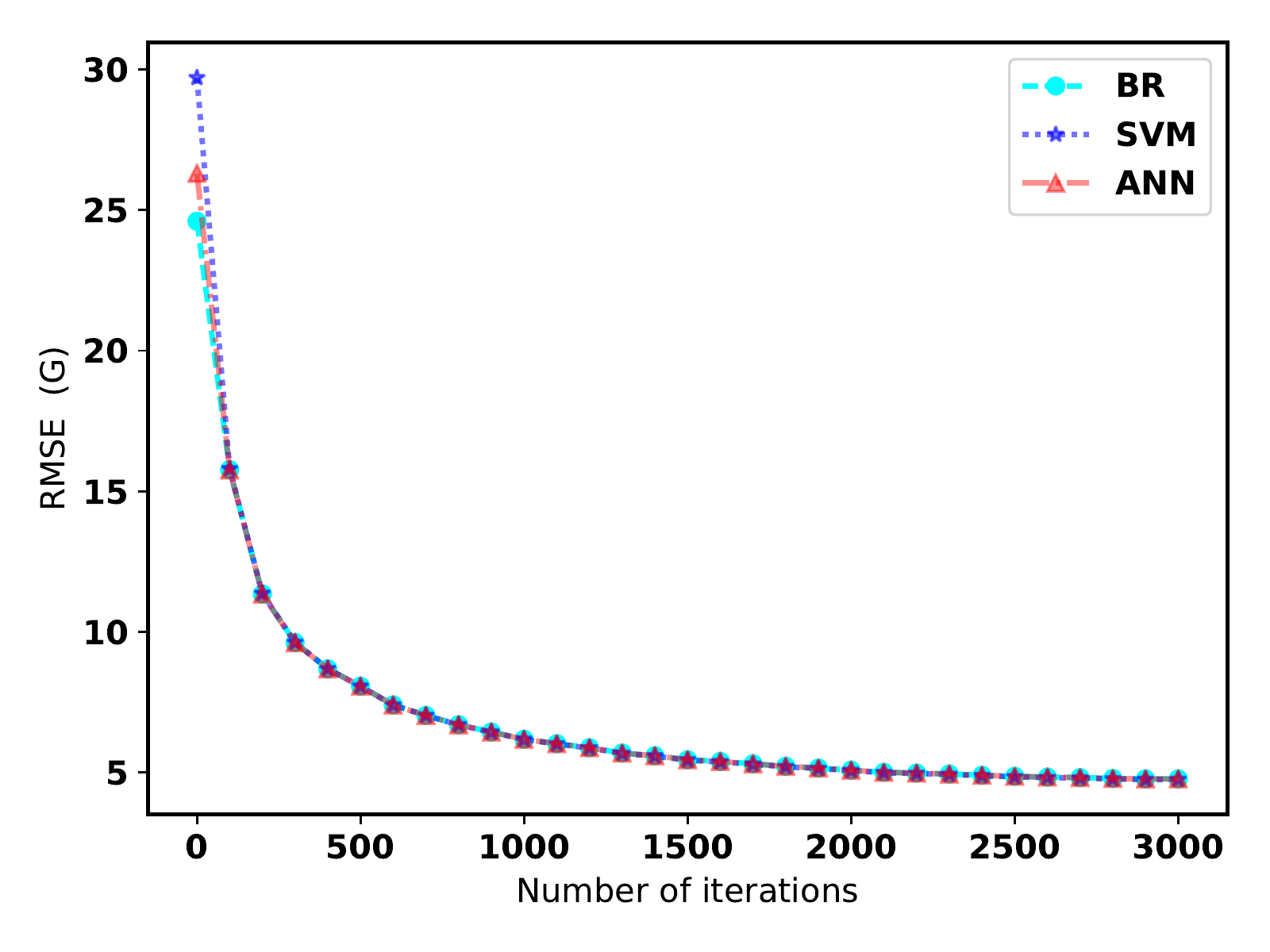}
\caption{Performance comparison of the three regression models as function
of the number of iterations when considering noise level of 0.4. The performance
of the three regressor is almost identical.}
\label{fig:error_iter}
\end{center}
\end{figure}

In order to inspect more carefully the dependence of the results on the number
of iterations, we tested the proposed inversion process using the full set of $1\,500$ MZSs.
Additionally, it is interesting to compare the performance of the three regression models 
introduced above. We recall that we are considering the same noise level of 0.4
for all the iterations and for all the MZSs. 
The RMSE of the inversions for the full set of profiles as function 
of the number of iterations is shown in Fig. \ref{fig:error_iter}. 
All the three regressor models show very similar behaivors, 
with a considerable improvement of the results specially for the first hundreds iterations:
When no iteration is applied the RMSE are 24.6, 29.7 and 26.3 G for 
the BR, SVM, and ANN respectively, and after 100 iterations the RMSE
is reduced to 16 G for the three models, and  at 1000 iterations the RMSE is of only 6 G. 
After around one thousand iterations, the improvement of the inversions becomes restrained.
With the present results,  we consider that apply 3000 iterations is good enough for the 
addopted inversion strategy.


%
%
%

Now, concerning the incertitudes associated to our measurements of $H_{\rm eff}$,  in  principle,
it is possible obtain an error estimation based in the the histograms of Fig. \ref{fig:histos_examples}. 
Nevertheless, by doing it we would be highly 
overestimating the errors: The standard deviaton of the fitted Gaussian distribution for 
the bottom left panel is 160 G, but in fact the error is much more smaller, being of only 4.1 G. 
Simarly, an overestimation of the error is also obtained in other example of Fig. \ref{fig:histos_examples}, 
at bottom rigth panel, where the standard deviation of the fitted distribution is 13.8 G but the 
real error is of only 1.2 G. Thus, in order to obtain realistic estimations of the errors, 
we next procced to inspect  the inversions over the full sample of $1\,500$ MZS.

\subsection{Caracterizing the technique}


We will now show the performance of our proposed method considering differents 
noise levels; we tested the inversions using the same set of $1 \,500$ MZSs as before,
and in all cases we applied $3\, 000$ iterations. 
To quantify the accuracy in the results we have employed
the Mean Absolute Percentual Error (MAPE), defined as:

\begin{equation}
 MAPE=100 * \left( \, \frac{ |\, H_{\rm eff}^{original} - \, H_{\rm eff}^{inferred}\, | }{|\, H_{\rm eff}^{original} \,|} \right).
\end{equation}

Additionaly, in order to underline the importance of including a noise treatment in the 
analysis of spectropolarimetric data, which in  our case is done through the noise 
cleaning of the MZSs followed by the iterative inversion procedure, we also included 
the results when the profiles were inverted without any noise treatment, 
which we labelled as \emph{single} inversions.

\begin{figure}
\begin{center}
 \includegraphics[width=8cm]{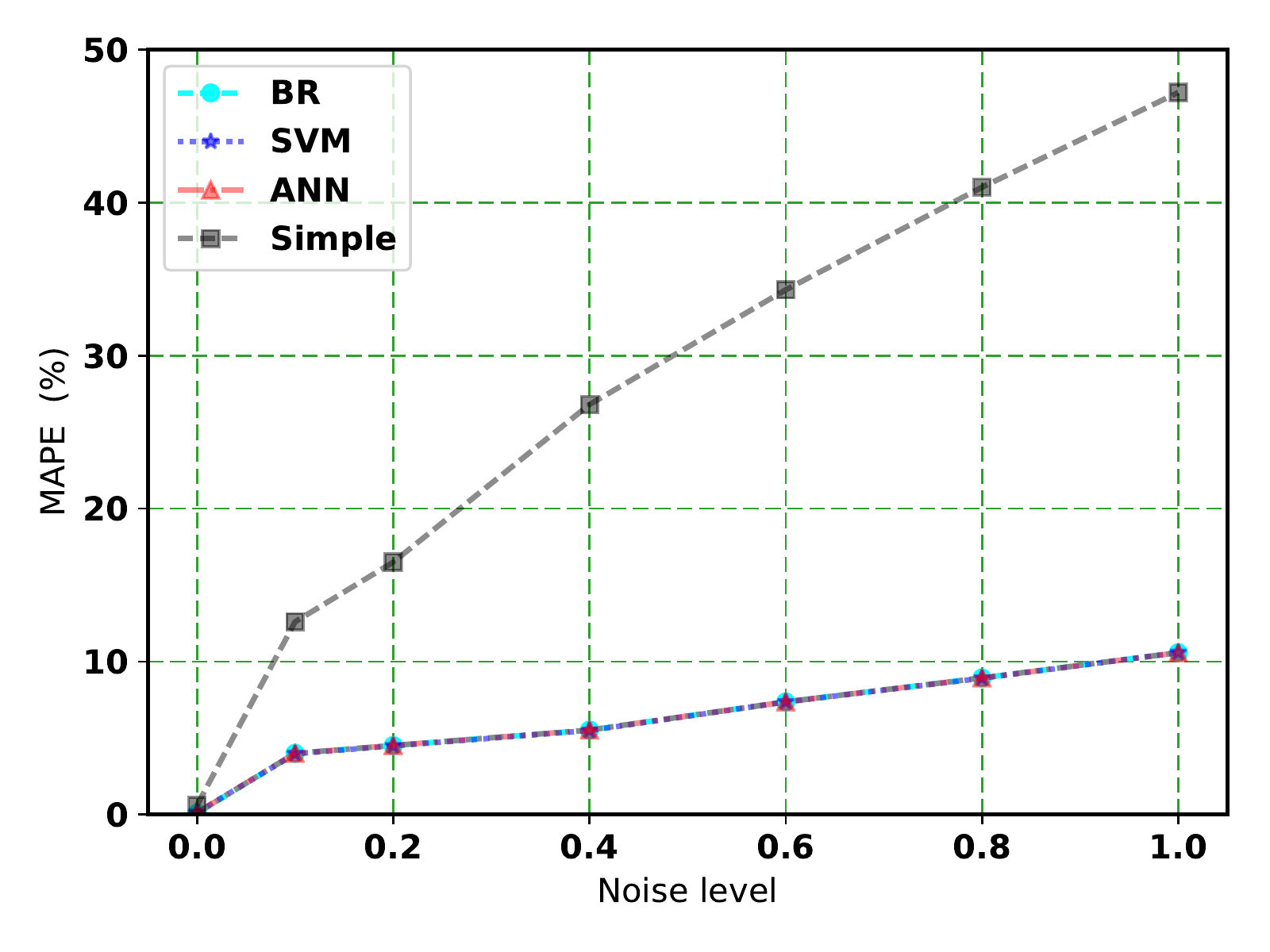}
\caption{Response of the MAPE of the inversions
considering differents noise levels.
Each of the three regression models, as well as the case of 
\emph{single} inversions, are indicated in the inner legends.}
\label{fig:mape_rmse}
\end{center}
\end{figure}

In Fig. \ref{fig:mape_rmse}, we show the variation of MAPE
as function of the noise level. The three regressor models
(BR, SVM, ANNN) have the same performance as function of the noise level:
When no noise is included the MAPE is extremely low (0.1\%), 
but the inclusion of even a very small noise level of 0.1  lead 
to a (relative) considerable increament of the MAPE (4\%). 
However, the precision of the inversions for the  models 
will then only decrease moderately to finish with a MAPE of 11\% for the highest 
noise level of 1. Finally, the case of traditional inversions --noted \emph{simple} 
by the line with square symbols--, is the most affected one, varying from 13\% for the 
lowest noise level of 0.1 up to  47\%  for the highest noise level of 1.

One conclusion from this test is that with our proposed
methodology -PCA noise cleaning combined with iterative
inversions-, the three regression models are robust methods to
determine $H_{\rm eff}$ from MZSs that are affected by the noise. 
Another important conclusion is that the {\em single} inversions
are highly affected by noise, being the respective errors 
much higher than those obtained with any of the regression models.
This is particularly evident for the case of highest noise level 
of unity, where the MAPE of the {\em single} inversions can reach 
almost 50\%, while for all three regressors, for this same noise level, 
the MAPE is close to 10\%.

Similar results have been previously presented in  \cite{carroll2014}.
In this work the noise treatment is done adopting a {\em sparsity}
representation of the data using an Ortogonal Match Pursuit (OMP) algorithm.
In Fig. 10 of that article, the authors presented an analogous test to the 
one just discussed here, comparing the inversions of noise affected profiles
using the OMP approach with the ones obtained using the traditional 
centre-of-gravity  (COG) method. The latter, is nowadays
the usual method employed when measuring  $H_{\rm eff}$ from real data and it
is relevant in the sense that it does not consider any noise treatment. 
The authors have found that when considering a noise level of 1,
using the COG method the MAPE is close to 50\% (the same as in our {\em single} case), 
while using the OMP approach the MAPE decreases to less than 20\%.  
Both results, theirs and ours, stand out the importance of implementing a noise 
treatment in the inversion of polarised multi-line profiles.

The iterative process described in this section is a time-expensive step
in our approach since it takes 3.75 hrs using  8 processors in parallel. Nevertheless
this relative long time is requiered in order to achieve the precision errors showed
in Fig.  \ref{fig:mape_rmse}, and of course this time can be reduced if more processors
are at disposse.

\section{Applying the method to real data}

\subsection{HD 190771}
In this section we will apply our method to real data. We have obtained from 
the public database {\em PolarBase}, 
spectropolarimetric data of two cool solar like stars, with
very similar physical properties to the Sun.
The first one is HD 190771, whose atmospheric parameters are:  $T_{\rm eff} = 5834 \pm 50$ K, 
log $g = 4.47 \pm 0.03$, $M = 0.96 \pm 0.13 \, M_{\sun}$, ${\rm M/H}=0.14$ and 
$vsini  =  4.3 \, \rm km\,s^{-1}$ \citep{valenti2005}. 
For the synthesis of the stellar spectra, we took as starting point the closest model of 
the grid of Atlas9 atmospheric models ($T_{\rm eff} = 5750 $ K, log $g = 4.5 $ and  ${\rm M/H}=0.0$),
to then extrapolate to the exact values of $T_{\rm eff}$, log $g$  and 
metallicity \citep[see for example,][]{castelli2005}.

\begin{figure}[ht]
\begin{center}
 \includegraphics[width=8cm]{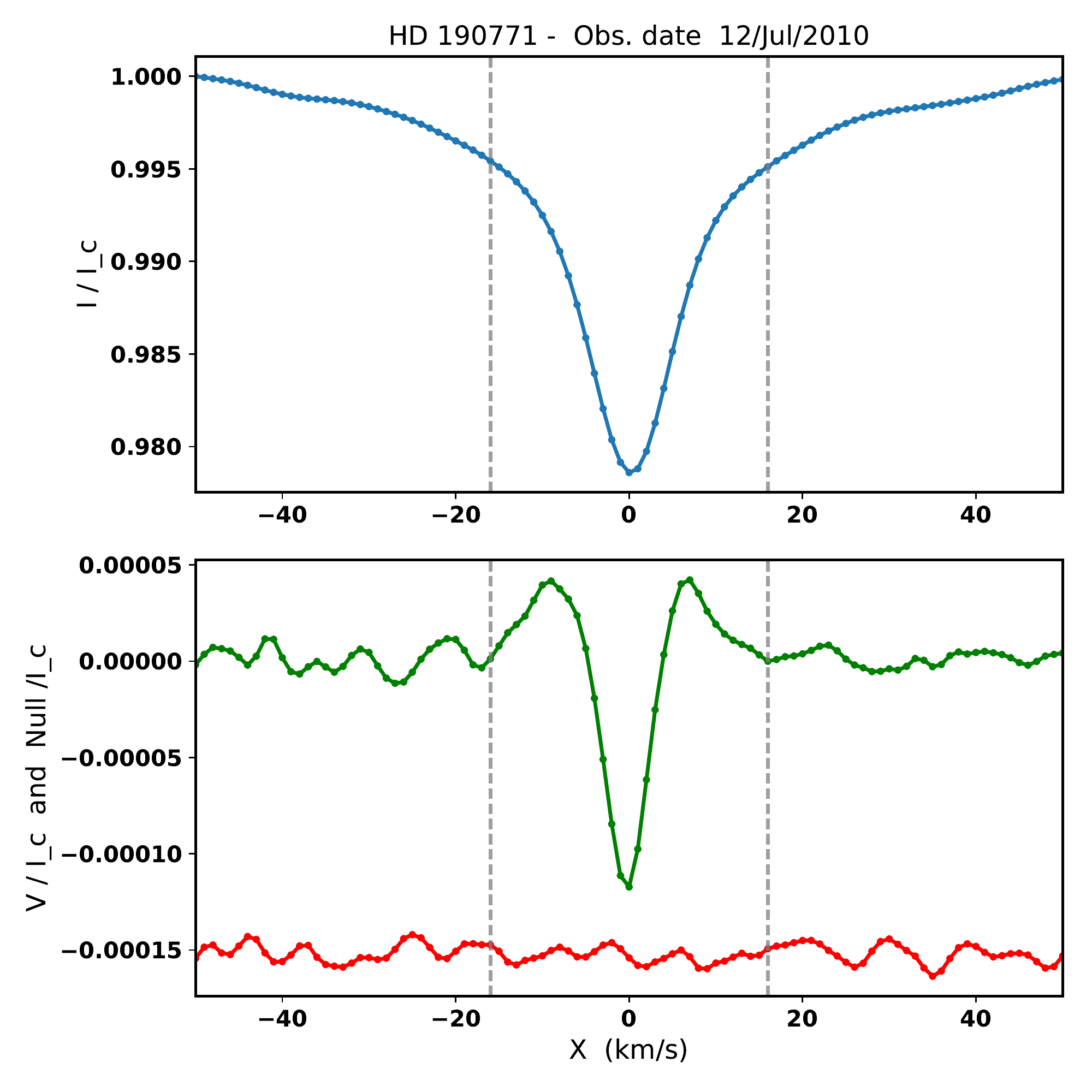}
\caption{From top to bottom, MZS profiles in intensity, circular polarisation 
and null in the reference frame of the star.  The central region, determined by the 
vertical lines, defines the rotational span velocities for this star in the doppler
space, from -16 to 16 ${\rm km\,s^{-1}}$ .}
\label{fig:mzs_hd190991}
\end{center}
\end{figure}

The synthesis of the spectra covered a wavelength range of 369-1010 nm in steps
of 1 ${\rm km\,s^{-1} }$. 
As mentioned in \citetalias{jcrv2016}, when constructing the MZSs 
our only criteria for the line selection  is a minimum ratio of line-depth to continuum, 
which for this work we have fixed > 0.1, resulting in a total number of individual
lines of $23\,172$.
The established MZS for the intensity, circular polarisation 
and null profiles are shown Fig. \ref{fig:mzs_hd190991}.

\begin{figure*}[h]
\begin{center}
 \includegraphics[width=16cm]{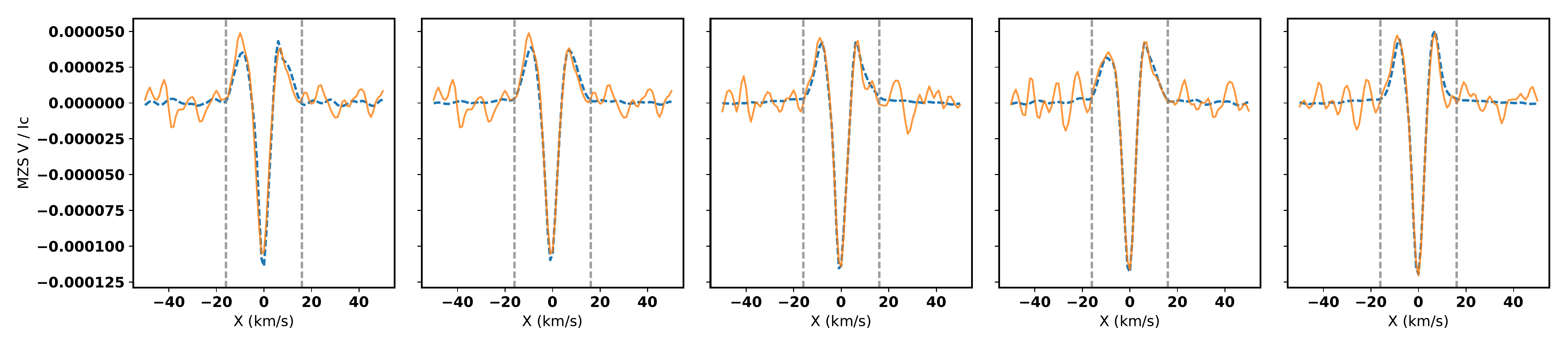}
\caption{Examples of MZS profiles obtained for the HD 190771 data (solid lines) 
and the respective  denoised MZSs (dashed lines).
The shape varitions of the MZSs are similar to those in the synthetic case showed 
in the upper panel of Fig. \ref{fig:cleaned_mzs_iter}. The dashed vertical lines 
determine the region employed in the inversion of the MZS profiles.}
\label{fig:hd190771_mzs_iter}
\end{center}
\end{figure*}

\begin{figure}[h]
\begin{center}
 \includegraphics[width=7cm]{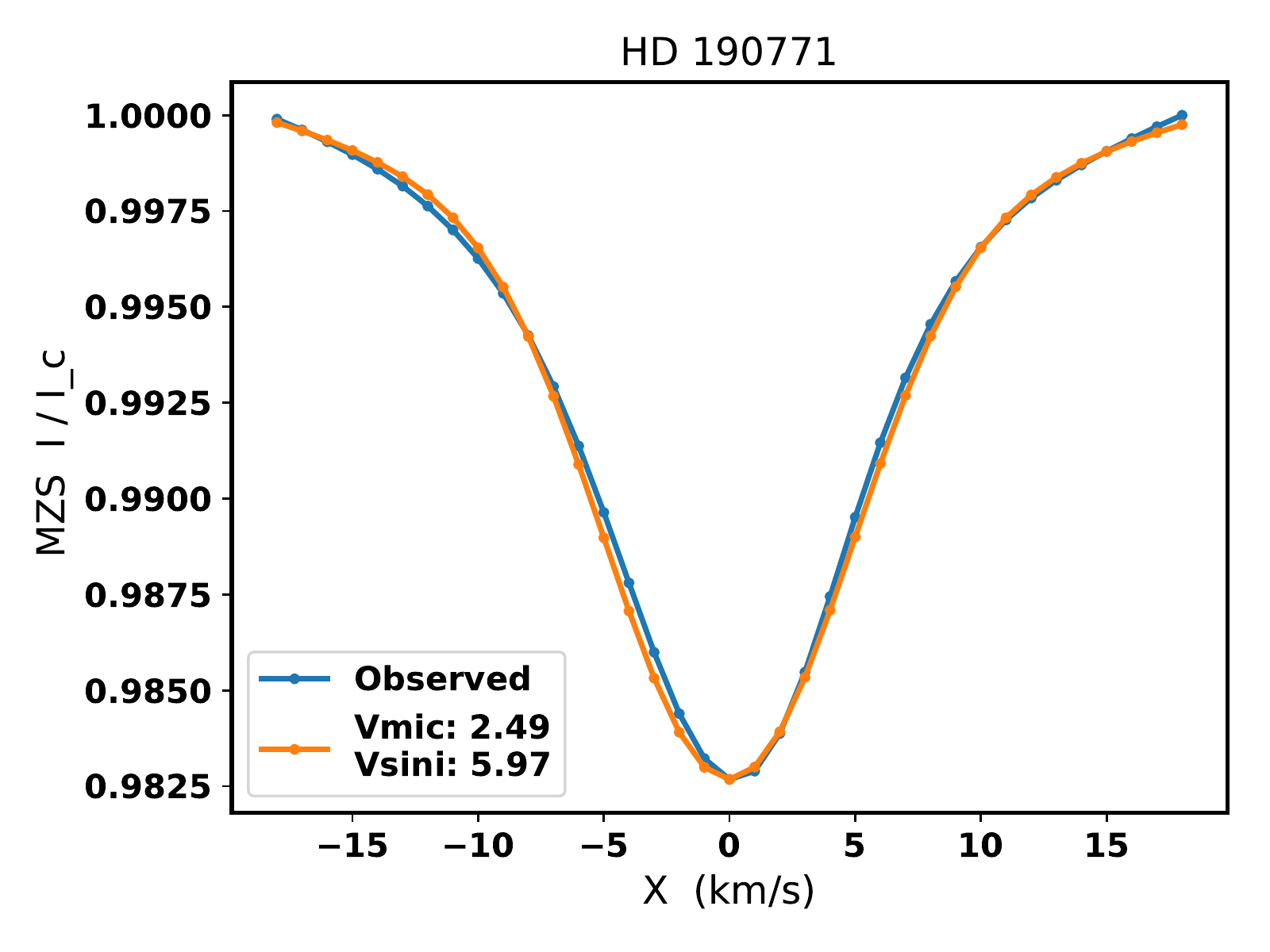}
\caption{MZSs in intensity for the data and the best fit model considering as 
free parameters $vsini$ and $\xi$ (both in units of ${\rm km\,s^{-1}}$).}
\label{fig:mzs_i}
\end{center}
\end{figure}

The next step is to fit the observed broadening for the intensity Stokes profile. 
For this propose, we set as free parameters the projected rotational velocity of 
the star ($vsini$) and the microturbulence velocity  ($\xi$):  
We calculated a sample of 50 stellar spectra varying $vsini$=[0.5,10] and $\xi$=[0,5]
${\rm km\,s^{-1}}$. We then  trained an algorithm to predict simoultaneously 
both parameters, obtaining as  best fit values $vsini=5.97$ and $\xi=2.49$ ${\rm km\,s^{-1}}$.
In Fig. \ref{fig:mzs_i} we show the obtained fit of 
synthetic profile to the observed one. 
In the adjustement of the intensity  MZS profile, we are considering that all the other 
broadening processes such as the magnetic, instrumental or macroturbulent ones, can 
be well represented by the interplay between only these two free parameters.
For this reason, the values of
$vsini$  and $\xi$ derived from this fitting are not real \emph{measurements} of these
two physical parameters.

In summary, we use an optimal atmospheric model ($T_{\rm eff}=5834$ K, 
log $g=4.47$, ${\rm M/H}=0.14$, $vsini=5.97$ ${\rm km\,s^{-1}}$ and 
$\xi=2.49$ ${\rm km\,s^{-1}}$)\, to synthesize a set of 300
MZS that we will use to train the regression algorithms and to subsequently
determine $H_{\rm eff}$ for this star.

In order to follow a consistent MZSs inversion procedure
with our previous tests, it is required to determine the standar deviation of both, 
the noise-free MZS ($\sigma_{MZS}$) and the noise ($\sigma_{noi}$).
The latter was determined using the edges of the circular MZS profile 
--indicated by the dashed vertical lines in Fig. \ref{fig:mzs_hd190991}--, 
because in these regions there is only noise.
For the former, we considered only the central region 
where is present the polarised signature. 
We then applied  a noise cleaning process to the polarised MZS as described in 
the precedent section. Once we have obtained the cleaned MZS, we calculated
the standard deviation from this noise-free profile. Finally, the noise level is
determined as before, NL = $ \sigma_{noi} \, / \sigma_{MZS}$, obtaining a 
value of 0.15.

\begin{figure}[t]
\begin{center}
\includegraphics[width=8cm]{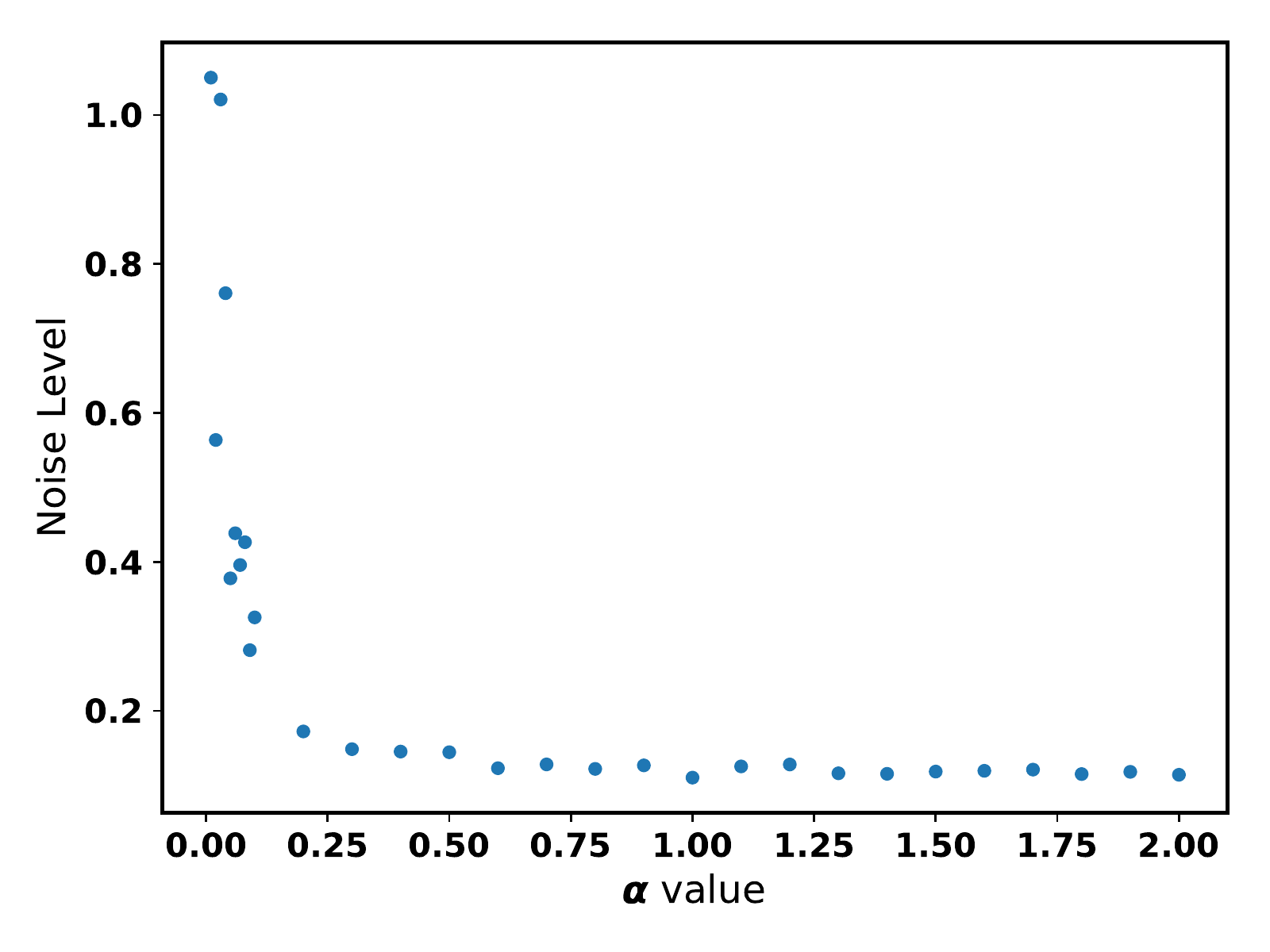}
\caption{Variation of the noise level of the MZSs as function of the 
$\alpha$ value used in Ec. (\ref{ec:A_i}); see text for details.}
\label{fig:alpha_NL}
\end{center}
\end{figure}

In order to apply the iterative inversions procedure, in each of the 39 
echelle orders we added random white noise to the observed circular polarised 
spectra. The amplitude ($A_i$) of the added noise in each order is different,
and it is given by :

\begin{equation}
A_i = \sigma_{i}^{\rm v} / \alpha,
\label{ec:A_i}
\end{equation}
where $\sigma_{i}^{\rm v}$ is the standard deviation of the circular polarised spectra 
in  the i-order, and $\alpha$ is a constant factor for all orders. It is through the
value of $\alpha$ that the NL can be controlled in the MZS after the 
addition of random white noise. In Fig. \ref{fig:alpha_NL}
we show how the noise level of the MZSs varies as function of the $\alpha$ value
for this particular observation of HD 190771: 
For very low values of  $\alpha$ --corresponding to high amplitudes in the added noise--,  
the NL is very high,  but as $\alpha$ increases the NL decays very fast, to finally 
become stable  at around  NL $\sim$ 0.15, which correspond to the noise level 
of the MZS when no noise was added. 
In this plateau phase of the NL, corresponding to $\alpha$ values from 0.3 to 2, 
we are interested in the  minimum ($\alpha$ = 0.3), because the lower is the value of $\alpha$ 
the higher will be the added noise, i.e, the higher will be the variations in the 
iterative establishment of MZS profiles. 
In Fig. \ref{fig:hd190771_mzs_iter}, we show some examples of the of MZS profiles 
for the HD 190771 data considerig $\alpha=0.3$.

\begin{figure}[h]
\begin{center}
 \includegraphics[width=8cm]{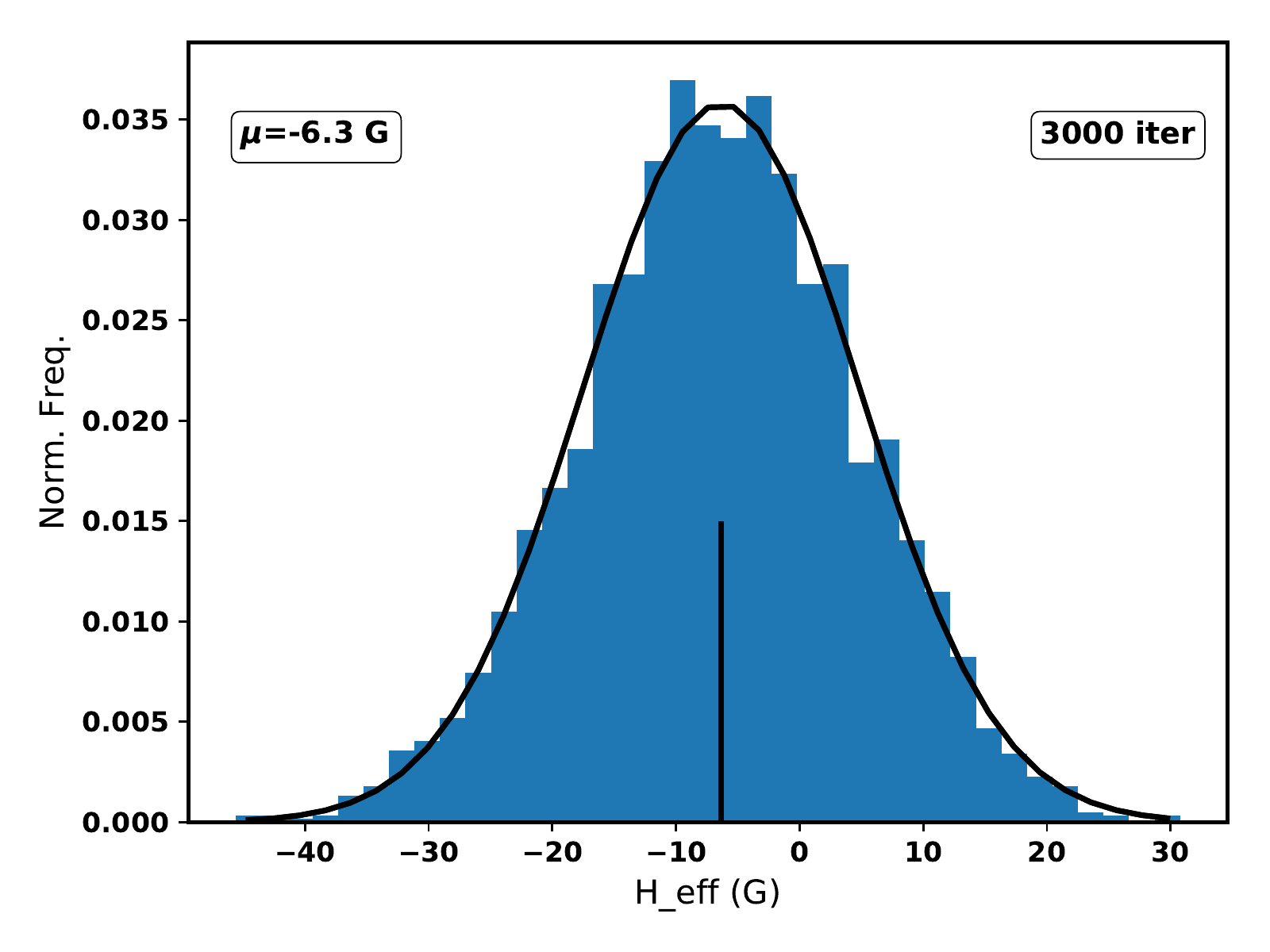}
\caption{Histogram of the inversions profiles for the star HD 190771, from 
we determined a value of $H_{\rm eff} = -6.3$ G. The regressor algorithm employed
in the inversions is the BR.}
\label{fig:histo_hd190771}
\end{center}
\end{figure}

Finally,  we considered $3\, 000$ iterations to produce the histogram of the inversions results.
In Fig. \ref{fig:histo_hd190771} we show the distribution of the inversions,  as well as the fit 
of the normal Gaussian distribution, from which we derived a value of  
$H_{\rm eff} = -6.3 \pm 0.3 $ G. Using the same dataset, \cite{marsden2014}
have reported a value of $H_{\rm eff} = -9.8 \pm 0.3$ G.

The value of the error in our measurement of $H_{\rm eff}$ 
is based on the MAPE results of Fig. \ref{fig:mape_rmse}. 
Considering that in our inversions we employed the BR regressor and
given that the  noise level of the data is 0.15,   
it corresponds to a MAPE close to 5\%, i.e., an error of $\pm$ 0.3 G. 
It is worth noticing  that the same 
error is reported by  \cite{marsden2014} despite the fact that in the inversion of the LSD 
profile, the authors they do not implement any noise treatment. As a reference,
for the same NL of 0.15, the MAPE associated to \emph{single} case 
(inversions without noise treatment) is 15\%, i.e., 
the error should be around $\pm$ 1.5 G. It seems thus  that the errors
reported by classical techniques that do not consider noise traetment could be 
seriously underestimated.

Now, if we consider the estimation that we found of 1.5 G for the error  in the 
LSD measurement of $H_{\rm eff}$,  then it turns out that the value of  $H_{\rm eff}$ reported by 
\cite{marsden2014} and the one we found,  would be in agreement at a difference of
2$\sigma$. On the contrary, if we consider the error of 0.3 G reported by the authors, then 
the difference could not be explained by the incertitudes associated to the measurements, 
and in this case the discrepance in the results deserve
a much deeper analysis, which is beyond the scope of the present study.

\subsection{HD 9472}

\begin{figure}[b]
\begin{center}
 \includegraphics[width=8cm]{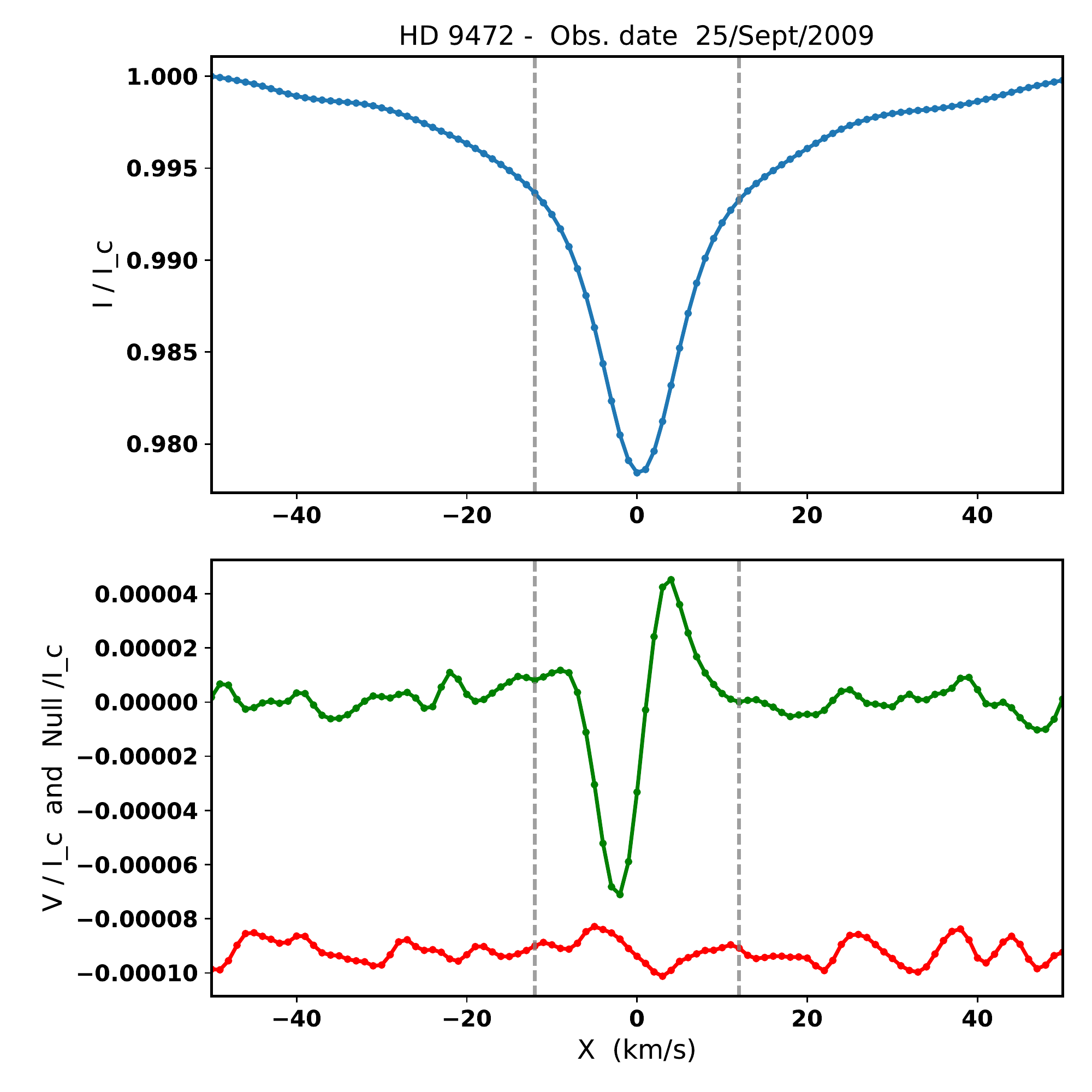}
\caption{Same as Fig. \ref{fig:mzs_hd190991}, but for HD 9472. The star spans in the doppler space
from -12 to 12 ${\rm km\,s^{-1}}$.}
\label{fig:chd9472}
\end{center}
\end{figure}

The second example is the star HD 9472, whose atmospheric parameters are
$T_{\rm eff}=5867 \pm 44$ K, log $g=4.67 \pm 0.02$, 
$M=1.29 \pm 0.19 \, M_{\sun}$, ${\rm M/H}=0.0$ and  $vsini=2.2 \,
{\rm km\,s^{-1}}$ \citep{valenti2005,marsden2014}.

We have followed the same procedure as in the previous case, taking as point of departure
the closest atmospheric model from the Atlas9 grid, namely, $T_{\rm eff} = 5750 $ K, 
log $g = 4.5 $, and  ${\rm M/H}=0.0$, to then extrapolate to the exact 
values of $T_{\rm eff}$ and  log $g$. Adopting the same threshold of the line-depth to continuum as before (> 0.1), 
the total number of individual lines is $23\,895$ for this star.  We then established the MZSs,
shown in Fig. \ref{fig:chd9472}.

\begin{figure}[h]
\begin{center}
 \includegraphics[width=7cm]{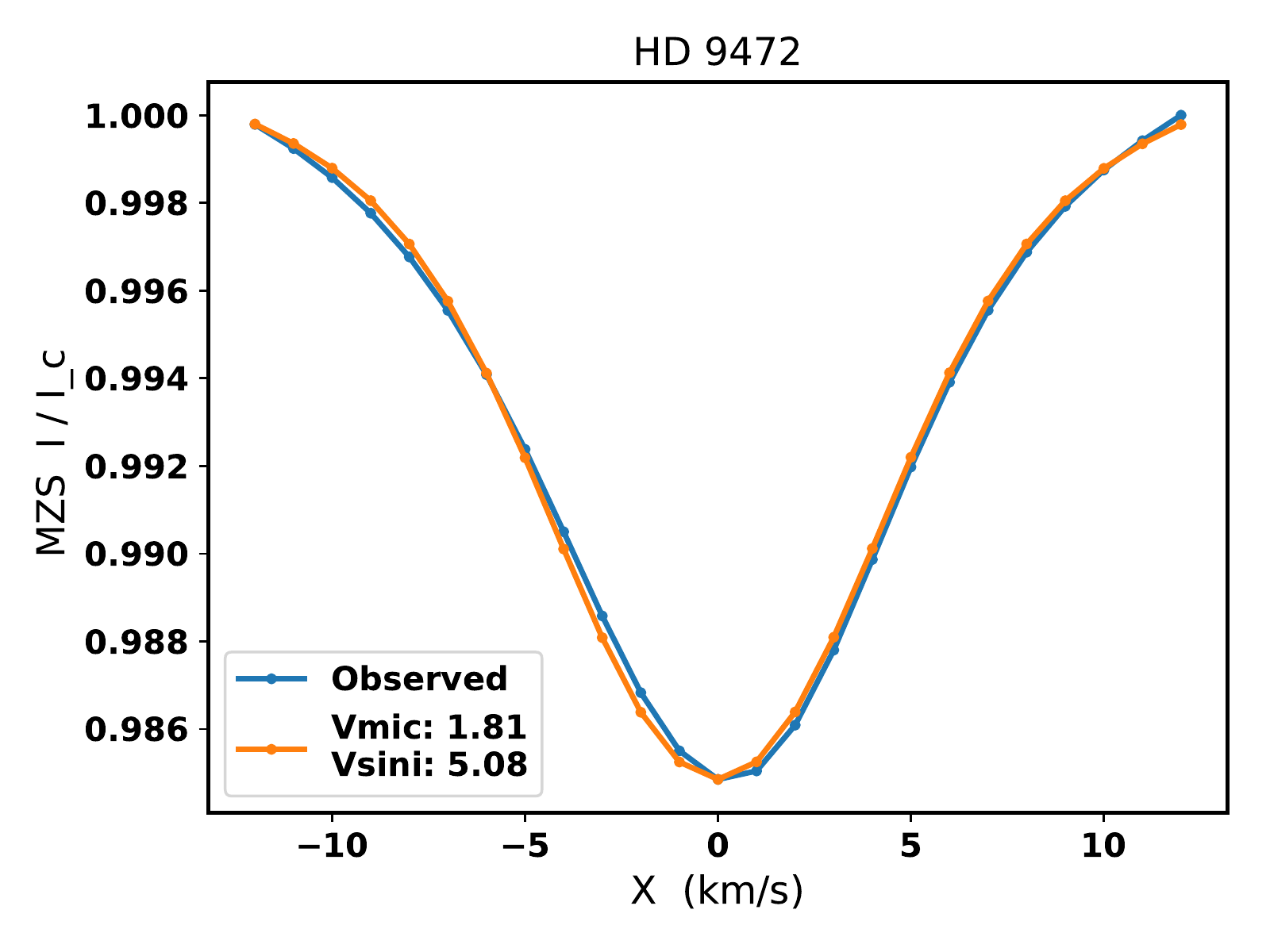}
\caption{Same as Fig. \ref{fig:mzs_i}, but for HD 9472.}
\label{fig:fit_mzsI_hd9472}
\end{center}
\end{figure}

The next step was to reproduce the observed broadening in the intensity Stokes MZSs.
We obtained as best fit values $vsini=5.08$ and $\xi=1.81$ ${\rm km\,s^{-1}}$,
which produced, as in the precedent case, a very good fit between the synthetic and
the observed profiles, as showed in  Fig. \ref{fig:fit_mzsI_hd9472}.

\begin{figure}[h]
\begin{center}
 \includegraphics[width=8cm]{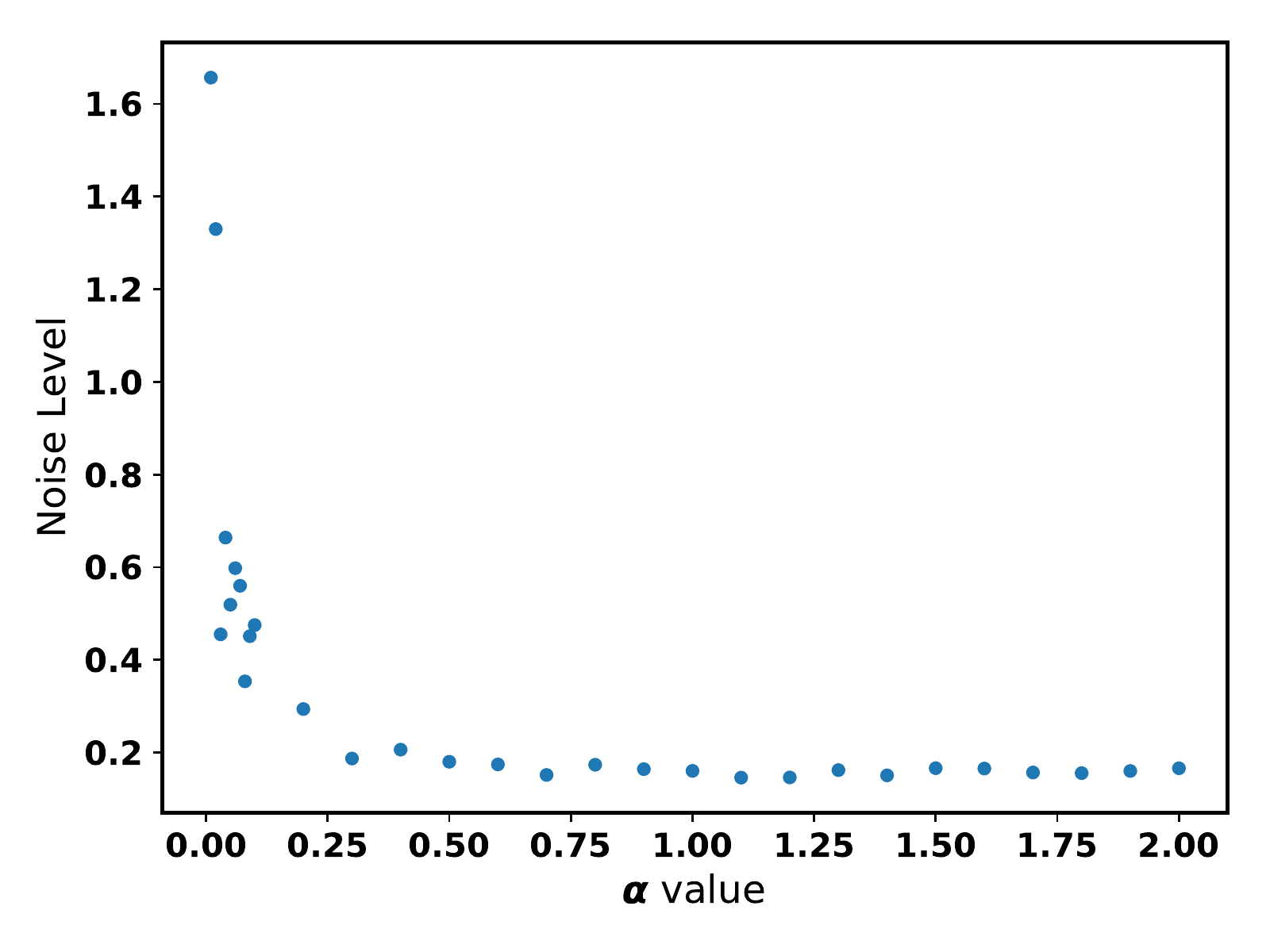}
\caption{Same as Fig. \ref{fig:alpha_NL}) but for the data of HD 9472.}
\label{fig:alpha_NL_hd9471}
\end{center}
\end{figure}

\begin{figure}[h]
\begin{center}
 \includegraphics[width=8cm]{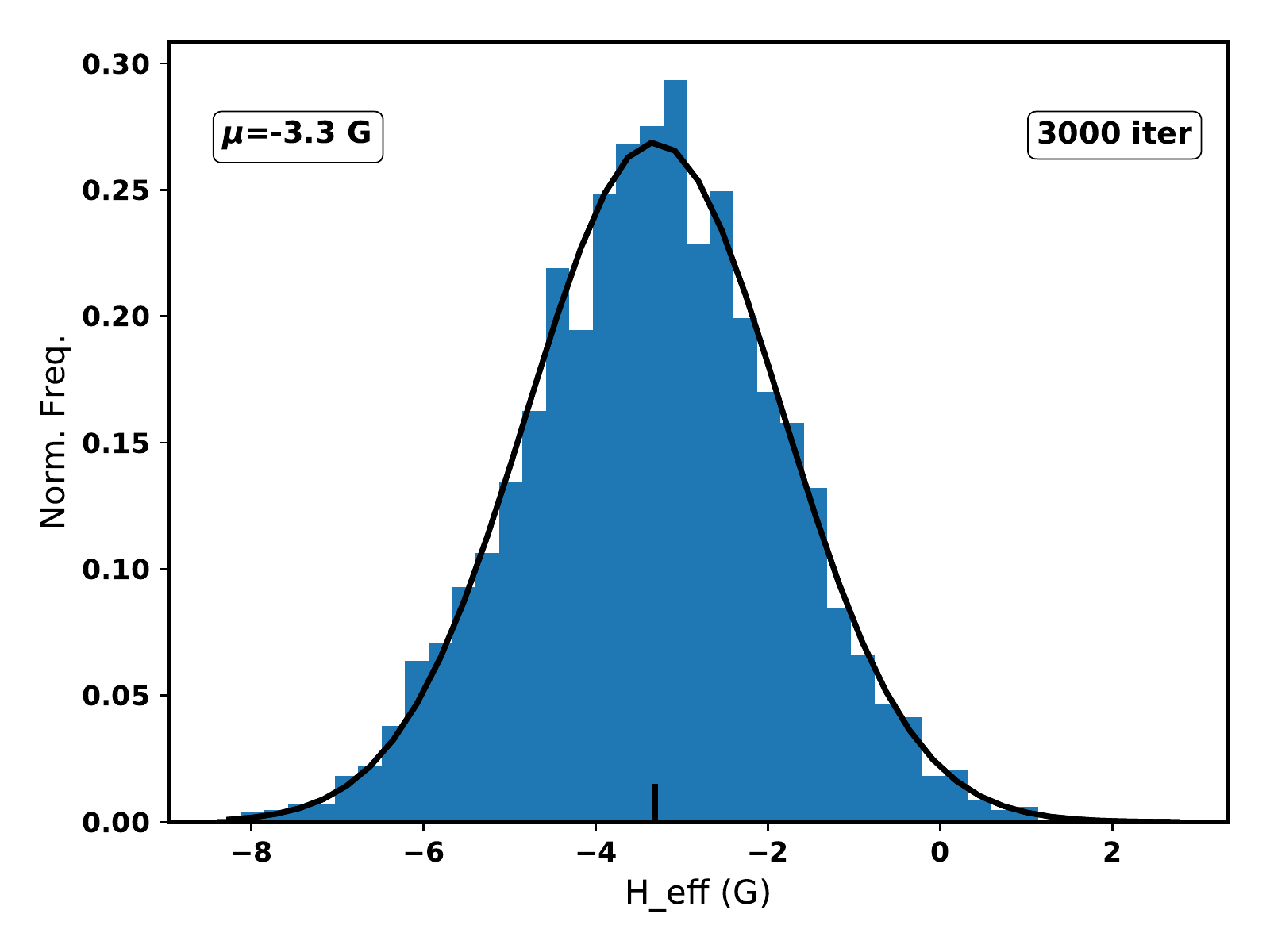}
\caption{Histogram of the inversions profiles for the star HD 9472, from 
we determined a value of $H_{\rm eff} = -3.3$ G. The regressor algorithm employed
in the inversions is the BR.}
\label{fig:histo_hd9471}
\end{center}
\end{figure}

The NL was then calculated  for the circular MZS, obtaining a
value of 0.19.  We subsequently inspected the variation of the noise level as function
of the amplitude of the random noise added to the echelle orders, 
allowing to determine that setting $\alpha =0.4$ will correspond
to MZSs with the same NL as the original data of 0.19, see Fig. \ref{fig:alpha_NL_hd9471}. 
Finally, we  iteratively stablished $3\, 000$ MZSs 
to consequently fit a Gaussian distribution to the histogram of the inversions,  
shown in Fig. \ref{fig:histo_hd9471}. The centroid of the fitted distribution 
corresponds to a value of $H_{\rm eff} = -3.3 \pm 0.2$ G, which in fact 
is almost the same that the one reported  by  \cite{marsden2014} using the same data:
$H_{\rm eff} = -3.5 \pm 0.4$ G. Once more, it seems that the error
reported by the authors is underestimated: based in Fig. \ref{fig:mape_rmse},
for this noise level the error should be around 17\%, i.e., 0.6 G.

\section{Conclusions}

We have presented a new study related to the measurement of stellar 
longitudinal magnetic fields, from high resolution spectropolarimetric data,
analysing multi-line profiles. 
Our main goal was to develope a general method which is constrained nor by 
the regime of validity of the weak field approximation
neither by the line autosimilarity normally assumed in the classical 
methods employed nowadays in the measurement of $H_{\rm eff}$. 

Our technique is thus based on a theoretical radiative transfer approach where  
we produced a synthetic set of Stokes profiles, each one corresponding to a different
configuration of the magnetic fields over the stellar surface. Considering the best 
possible scenario of noise-free spectra, we have showed that the 
use of machine learning algorithms (MLA) is a key step to reduce the number of 
synthetic profiles required for an acceptable accuracy in the correct 
determination of $H_{\rm eff}$: With the use of MLA and 
considering only 50 stellar spectra, we obtained a better value of the RMSE for the inversion
results that when we considered $7\, 500$ spectra but the inversion strategy was based 
in a look-up table, i.e.,  no MLA was applied \citep{jcrv2016}.

We also showed that the results are considerably degradated when noise is included; 
nevertheless this is not restricted to the MLA but in general to any method dealing with 
the inversion of noise-affected profiles or signals. We thus proposed a data noise pre-processing
consisting of two steps: 1) To use PCA to perform a noise-cleaning of the multi-line  profiles 
and 2) An iterative inversion procedure.  Applying this data analysis process, we achieved a considerable
improvement in the inversion accuracy, confirming that it is desirable  to include a 
noise treatment  in the analysis of multi-line profiles in order 
to get more confidence in the derived values of $H_{\rm eff}$. Very similar results
about the  impact of the noise  in the inversion of multi-lines profiles were independently found 
by \cite{carroll2014}.

We implemented our inversion technique to real observations of two stars. In order to
similarly apply the MLAs to the observed data set (i.e. as closely as possible
as in the case of synthetic spectra),  
it was first required to reproduce 
the observed broadening in the intensity profile. We allowed the variation of
two physical parameters for this propose, namely, $vsini$ and $\xi$, and 
we obtained very good fits, as shown in Figs. \ref{fig:mzs_i}
 and \ref{fig:fit_mzsI_hd9472}. Of course, the  values of these two parameters
lack of any physical interpretation because we are assuming that all other
broadening mechanisms (e.g. magnetic, instrumental or any other) can be well
represented by considerig only $vsini$ and $\xi$ as adjustable parameters.

Thus, the atmospheric model employed for the analysis of each star consisted in  
the exact values of $T_{\rm eff}$, log $g$ and metalicity --reported 
by other authors--, in addition to the values of $vsini$ and $\xi$. 
Once these five atmospheric parameters were defined, we synthesized a set 
of 300 polarised stellar spectra that in turn were used to train the MLA
(the used training data sets can be found in \url{www.astrosen.unam.mx/\~julio/ML_mzs}).
One final step was to reproduce the iterative inversions methodology implemented in the
synthetic tests. For this purpose, we added random white noise in each 
order of the echelle spectrum. The amplitude of the added noise in each order 
is different and is controlled by one free paramter, labelled $\alpha$ in Eq. (\ref{ec:A_i}). 
It is through the value of this parameter that it was possible to keep a constant noise level 
in  the MZSs established for the iterative inversions. Finally, considering the same 
number of iterations in the case of real 
data than in the tests,  we fitted  a Gaussian normal function to the distribution
of the inversion results. The centroid of the Gaussian distribution  determines the value of 
$H_{\rm eff}$ for each data set (Figs. \ref{fig:histo_hd190771} and \ref{fig:histo_hd9471}).

We have used two stars as test cases to illustrate the direct application of 
the proposed methodology to real 
observations, and in this sense,  it is not the scope of this work
to compare the measured values of $H_{\rm eff}$ with our technique
with the ones obtained by other methods. We will proceed to the comparison 
of the values of  $H_{\rm eff}$ obtained with classical methods versus ours 
in a forthcoming article.
However, what we can highlight the fact that the uncertainties
reported in the measurement of $H_{\rm eff}$ form the LSD profiles are most likely
underestimated. 
\cite{asensio_petit2015} have recently presented a method to establish
and analyse LSD profiles under a Bayessian framework. One of the
advantages of this approach is that it can estimate the intervals of credibility 
at each velocity point of the LSD profiles. Unfortunately, 
although the authors applied their method to three different stars, 
they did not report the respective measurements of $H_{\rm eff}$, preventing
the comparaison of their estimation of errors with other techniques.
As far as we know, no other work has addressed a detailed study 
on the expected uncertainties associated to the measurements of $H_{\rm eff}$ 
from noise-affected LSD profiles.

The most relevant aspect of the work presented here is not that 
we have included estimations of the uncertainties in the measurements
of $H_{\rm eff}$, but the fact that we have introduced a new 
methodology for the analysis of polarised multi-line profiles
--including a radiative transfer theoretical framework in the synthesis
of the Stokes profiles--, 
using a strategy based on machine learning algorithms. 
We  expect  to apply our data analysis method to stars other 
than solar-type stars in incoming studies.

\section*{Acknowledgements}
We are very greatful to the whole group of {\em Laboratorio de C\'omputo Inteligente del CIC-IPN
\footnote{http://www.cic.ipn.mx/sitioCIC/index.php/pre-rncnc}} 
for very constructive and illustrative discussions
about machine learning methods. We would also like to express our acknowledgement to
the anonymous referee for all the comments that helped to improve the clarity of
this work.
JRV acknowledges support grants from {\sc CONACyT 240441} and
{\sc Supercomputo – LANCAD-UNAM-DGTIC-326}.
Thanks go to AdaCore for providing the GNAT GPL Edition of its Ada compiler.

\bibliographystyle{aa}
\bibliography{ref_ml_mzs}
\begin{appendix}
\section{Hyperparameters for each regressor}
For the definition of each of the hyperparameters listed bellow, we refer the reader 
to \emph{Scikit-learn} user guide. For consistence, we labeled each  parameter 
following the mentioned guide 
\footnote{http://scikit-learn.org/stable/\_downloads/scikit-learn-docs.pdf}.
\subsection{Bayessian Ridge}
BayesianRidge(alpha\_1=1e-06, alpha\_2=1e-06, compute\_score=False, copy\_X=True,
       fit\_intercept=True, lambda\_1=1e-06, lambda\_2=1e-06, n\_iter=300,
       normalize=False, tol=0.001, verbose=False)

\subsection{Artificial Neuronal Network}
MLPRegressor(activation='identity', alpha=0.05, batch\_size='auto', beta\_1=0.9,
       beta\_2=0.999, early\_stopping=False, epsilon=1e-08,
       hidden\_layer\_sizes=(100,), learning\_rate='constant',
       learning\_rate\_init=0.001, max\_iter=10000, momentum=0.9,
       nesterovs\_momentum=True, power\_t=0.5, random\_state=25, shuffle=True,
       solver='lbfgs', tol=1e-08, validation\_fraction=0.1, verbose=False,
       warm\_start=False)
       
\subsection{Support Vector Machine}
SVR(C=50.0, cache\_size=200, coef0=0.0, degree=3, epsilon=0.1, gamma='auto',
  kernel='linear', max\_iter=-1, shrinking=True, tol=0.001, verbose=False)

\end{appendix}
\end{document}